\documentclass[11pt,a4paper]{article}

\pdfoutput = 1

\usepackage{jcappub}
\usepackage{multirow}
\pdfoutput=1
\usepackage[utf8]{inputenc}
\usepackage{float}
\usepackage{hyperref}
\usepackage{amsmath}
\usepackage{graphicx}
\usepackage{dcolumn}
\usepackage{bm}
\usepackage{epsfig}
\usepackage{amssymb,latexsym,mathrsfs}
\usepackage{graphicx}
\usepackage{color}
\usepackage{booktabs}

\usepackage{comment}

\definecolor{myred}{RGB}{180,50,28}
\definecolor{myblue}{RGB}{2,50,180}
\definecolor{mygreen}{RGB}{2,150,80}

\hypersetup{colorlinks=true,
            citecolor=blue,
            linkcolor=blue,
            urlcolor=blue}

\newcommand{\be}{\begin{equation}}
\newcommand{\ee}{\end{equation}}
\newcommand{\bea}{\begin{eqnarray}}
\newcommand{\eea}{\end{eqnarray}}

\newcommand{\bra}{\alpha_{\textrm{B}}}
\newcommand{\run}{\alpha_{\textrm{M}}}
\newcommand{\kin}{\alpha_{\textrm{K}}}
\newcommand{\ten}{\alpha_{\textrm{T}}}

\newcommand{\lcdm}{$\Lambda$CDM}

\newcommand{\CLASS}{\textsc{class}}
\newcommand{\hiclass}{{\tt hi\_class}}

\newcommand{\Pnon}{P_{\mathrm{non}}}
\newcommand{\Plin}{P_{\mathrm{lin}}}

\newcommand{\Pnonref}{P^{\mathrm{ref}}_{\mathrm{non}}}
\newcommand{\Plinref}{P^{\mathrm{ref}}_{\mathrm{lin}}}

\newcommand{\Rnon}{R_{\mathrm{non}}}
\newcommand{\Rlin}{R_{\mathrm{lin}}}

\newcommand{\Bref}{B^{\mathrm{ref}}}

\title{Enabling matter power spectrum emulation in beyond-$\Lambda$CDM cosmologies with COLA}

\author[a,b,1]{Guilherme Brando,\note{Corresponding author.}}
\author[c]{Bartolomeo Fiorini,}
\author[c]{Kazuya Koyama,}
\author[d]{Hans A. Winther}

\affiliation[a]{Max Planck Institute for Gravitational Physics (Albert Einstein Institute) \\
Am Mühlenberg 1, 14476 Potsdam-Golm, Germany}
\affiliation[b]{PPGCosmo, CCE -- Universidade Federal do Esp\'irito Santo,\\
Avenida Fernando Ferrari
514, 29075-910 Vit\'oria, Esp\'irito Santo, Brazil}
\affiliation[c]{Institute of Cosmology and Gravitation, University of Portsmouth,\\ Dennis Sciama Building, Burnaby Road, Portsmouth PO1 3FX, United Kingdom}
\affiliation[d]{Institute of Theoretical Astrophysics, University of Oslo,\\ PO Box 1029, Blindern 0315, Oslo,
Norway}

\emailAdd{guilherme.brando@aei.mpg.de}

\abstract{We compare and validate COLA (COmoving Lagrangian Acceleration) simulations against existing emulators in the literature, namely Bacco and Euclid Emulator 2. Our analysis focuses on the non-linear response function, i.e., the ratio between the non-linear dark matter power spectrum in a given cosmology with respect to a pre-defined reference cosmology, which is chosen to be the Euclid Emulator 2 reference cosmology in this paper. We vary three cosmological parameters, the total matter density, the amplitude of the primordial scalar perturbations and the spectral index. 
By comparing the COLA non-linear response function with those computed from each emulator in the redshift range $0 \leq z \leq 3$, we find that the COLA method is in excellent agreement with the two emulators for scales up to $k \sim 1 \ h$/Mpc as long as the deviations of the matter power spectrum from the reference cosmology are not too large. 
We validate the implementation of massive neutrinos in our COLA simulations by varying the sum of neutrino masses to three different values, $0.0$ eV, $0.058$ eV and $0.15$ eV. 
We show that all three non-linear prescriptions used in this work agree at the $1\%$ level at $k \leq 1 \ h$/Mpc. We then introduce the Effective Field Theory of Dark Energy in our COLA simulations using the N-body gauge method. We consider two different modified gravity models in which the growth of structure is enhanced or suppressed at small scales, and show that the response function with respect to the change of modified gravity parameters depends weakly on cosmological parameters in these models.}
\begin{document}
\maketitle
\flushbottom

\section{Introduction} \label{intro}

One of the main efforts in Large Scale Structure (LSS) studies aims to answer how initial small and Gaussian density perturbations in the very early Universe evolved to the highly non-Gaussian and non-linear matter distribution we see in our Universe today. In the next years we will have the first data release of the first Stage-IV LSS survey, DESI~\cite{desi}, as well as the launch of the Euclid satellite~\cite{euclid}. The bulk of the investigation in recent years is focused on the exploration of the matter two point correlation function in Fourier space, the matter power spectrum. In order to get fast predictions of this statistics, emulation techniques have gained much attraction in cosmology, and they are now seen as viable alternatives for extracting parameter constraints using the data from upcoming surveys.  

Emulators are numerical interpolations that are trained using accurate N-body simulation outputs based on machine learning algorithms to quickly predict the matter power spectrum from linear to non-linear scales in the vast cosmological parameter space. Among the emulators already available in the literature, we will focus on two that have been considered as highly effective and validated, the Euclid Emulator 2\footnote{\href{https://github.com/miknab/EuclidEmulator2}{https://github.com/miknab/EuclidEmulator2}}~\cite{EE2} (EE2) and Bacco\footnote{\href{https://baccoemu.readthedocs.io/en/latest/}{https://baccoemu.readthedocs.io/en/latest/}}~\cite{bacco} in this paper, and use them to check the accuracy of predicting non-linear matter power spectra. Both have an accuracy of about $1\%$ on small scales for $\Lambda$CDM cosmologies, and around $3\%$ for dynamical dark energy and massive neutrinos cosmology in predicting the non-linear power spectrum. 

The process of training these emulators heavily relies on the use of computationally expensive and time-consuming full N-body simulations. To overcome these limitations, there are several well established methods that allow us to quickly generate approximate non-linear realizations of the matter density field, such as, the COLA (COmoving Lagrangian Acceleration) approach~\cite{tassev2013,tassev2015,lpicola}, EZMOCKS~\cite{ezmocks}, PATCHY~\cite{patchy}, FastPM~\cite{fastpm}, GLAM~\cite{glam}, to name a few. Specifically, in this work, we will consider the first of these examples, the COLA approach. This method has been well-validated, and is known to give a good agreement on quasi non-linear scales in $\Lambda$CDM and beyond-$\Lambda$CDM cosmologies when comparing its prediction for the matter power spectrum to the ones from full N-body simulations \cite{albert,Valogiannis:2016ane, bill0,bill,bart}. Additionally, a new avenue using the COLA method was presented in~\cite{necola}. In this paper, it was shown that the mapping from displacements in COLA simulations and those in full N-body simulations can be trained on simulations with a fixed value of the cosmological parameters, and this model can be used to correct the output of COLA simulations with different values of cosmological parameters including different masses of massive neutrinos with very high accuracy: the power spectrum and the cross-correlation coefficients are within $1\%$ down to $k=1 \ h/$Mpc. 

This indicates that the inaccuracy in COLA's predictions is fairly cosmological parameter independent, and this can be corrected by a cosmological parameter independent model. In this paper, we show that COLA is capable of describing the non-linear response of the matter power spectrum to the change of cosmological parameters down to $k=1 \ h/$Mpc with high accuracy. This is because the inaccuracy of COLA is largely cancelled by taking the ratio of power spectra in two different cosmologies. The accurate and fast computation of this response function is of great importance when building emulators, as it allows us to obtain quickly the expected non-linear prediction when parameters are varied with respect to a fixed pre-defined reference cosmology. Given the prediction of non-linear power spectra in a few sparsely sampled reference cosmologies by full N-body simulations, we can provide the prediction of the matter spectrum in a wide parameter space. This can be used to further extend the reach of the already accurate \lcdm \ emulators to beyond-\lcdm \ cosmologies, for example, where running full N-body simulations are very expensive. 

Within many different alternatives to the standard cosmology model, we will focus on studying the non-linear response function in scalar-tensor theories of gravity. These theories are possible explanations for the current accelerated expansion of our Universe. The simplest solution in General Relativity (GR) is the addition of the Cosmological Constant, $\Lambda$, to the Einstein's field equations. Modified gravity is an alternative to the addition of this constant, where the scalar field would naturally act as the driver of the late-time accelerated expansion. Therefore, creating new emulators, or extending current ones, to be able to quickly predict the small scale behavior of the matter power spectrum in these theories is of great importance for the upcoming LSS surveys. Examples of modified gravity emulators in the literature are~\cite{rama_mg,forge} where in both papers the emulators are built for one specific fully covariant model, $f(R)$.

In this work, we will use the model-independent approach of the Effective Field Theory of Dark Energy~\cite{eftofde1,eftofde2} (EFT of DE). Within this formalism, we can fully characterize the linear perturbation of Horndeski theories, using four time dependent functions. 
Under the quasi-static approximation~\cite{kk_mg}, the modification to the matter power spectrum is characterised by a modified Newton constant called $G_{\rm eff}$, which is a function of these four functions. In this work, to maintain the model-independent approach, we will assume small scale modifications are always characterised by this function in the modified Poisson equation. On large scales, we use the N-body gauge approach developed in~\cite{us1,us2} to turn our simulations fully relativistic so that the linear power spectrum in COLA simulations agrees with that predicted by the Boltzmann code. The advantage of this approach is that relativistic corrections can be included in the same way as massive neutrinos and there is no additional cost in including EFT of DE in COLA simulations compared with simulations with massive neutrinos in \lcdm \ models. Additional effects on small scales such as screening and baryonic effects~\cite{schneider,dai_2018} can be included separately, but these are highly model dependent, and we will not discuss these effects in this paper.

The outline of the paper is as follows. In Section~\ref{sec:method} we introduce the numerical tools we use in this work, our COLA implementation and the Bacco and EE2, and define the linear and non-linear response function as well as the boost function, which is the ratio between non-linear and linear power spectra. In Section~\ref{sec:LCDM} we present the results for \lcdm \ cosmology, varying cosmological parameters as well as the total mass of massive neutrinos. We then move to discuss the non-linear response function in modified gravity theories in Section~\ref{sec:MG}: in Section~\ref{sec:MG_fixed_cosmo} we present the results when considering a fixed cosmology while in Section~\ref{sec:MG_var_cosmo} we vary the values of standard cosmological parameters within a given modified gravity theory. We present our conclusions in Section~\ref{sec:concl}.

\section{Methodology}\label{sec:method}

\subsection{COLA}
In this section, we will present the methodology of each code used to model the non-linear power spectrum. We begin by discussing the quasi N-body code COLA, used to generate fast realizations of the non-linear cold dark matter density field. The COLA method relies on the use of second order Lagrangian perturbation theory (2LPT) and an N-body Particle-Mesh (PM) algorithm to simulate the non-linear structure formation of the Universe. The 2LPT part of the code is responsible to accurately model the Universe at large scales, while the PM part dominates the dynamics of cold dark matter and baryons (cdm+b) at mildly to non-linear scales. For our COLA simulations we have chosen to use a ``forward'' approach to initiate our simulations, that is, we have introduced the N-body gauge formalism~\cite{Nbody1} on the publicly available COLA-FML code\footnote{\href{https://github.com/HAWinther/FML}{https://github.com/HAWinther/FML}}. We compute using \hiclass~\cite{hiclass1,hiclass2} \ a linear density field, encoding the relativistic correction, which is added at each time-step of the simulation. Therefore, instead of using a back-scaling method, where the linear matter power spectrum is given at $z=0$ and then back-scaled to the initial redshift, $z_{\rm ini}$, of the simulation, we provide transfer functions as external data files from $z_{\rm ini}$ onto $z=0$. This has been shown~\cite{us1, us2} to ensure that the output of these simulations correctly describe the relativistic effects introduced by photons, neutrinos and dark energy at large scales, and at small scales we smoothly transit to the Newtonian description of gravity. We have used the same approach developed by~\cite{bill} to introduce relativistic corrections, but have made the modifications to accommodate N-body gauge quantities. The new Poisson equation solved by COLA up to the second order is the following:
\begin{align}\label{eq:COLA_impl}
   k^2 \Phi^{\rm GR} &= 4 \pi G_{\rm N} a^{2}\rho_{\rm m} \left( \delta_{\rm m}^{(1)} + \delta_{\rm m}^{(2)} \right) \nonumber \\
    &= 4 \pi G_{\rm N} a^{2} \left[ \rho_{\rm cb} \delta_{\rm cb}^{(1)} + \rho_{\rm GR}\delta_{\rm GR}^{(1)} + \rho_{\rm cb}\delta_{\rm cb}^{(2)} \right] \nonumber \\
    &= 4 \pi G_{\rm N} a^{2} \rho_{\rm m} \left[
      f_{\rm cb} \left(\delta_{\rm cb}^{(1)} + \delta_{\rm cb}^{(2)} \right) + f_{\rm GR} \delta_{\rm GR}^{(1)} \right].
\end{align}
where, 
\begin{align}
    \delta_{\rm GR} = \frac{\delta \rho_{\rm GR}}{\rho_{\rm GR}}, \ \ \ \ \rho_{\rm GR} = \rho_{\nu} + \rho_{\gamma} + \rho_{\rm ur}, \ \ \ \ f_{\rm cb} = \frac{\rho_{\rm cb}}{\rho_{\rm m}}, \ \ \ \ f_{\rm GR} = \frac{\rho_{\rm GR}}{\rho_{\rm m}}, 
\end{align}
$\delta_{\rm cb}$ is the density contrast of baryons and cold dark matter, and $\rho_{\rm cb}$, $\rho_{\nu}$, $\rho_{\gamma}$ and $\rho_{\rm ur}$ are the background energy density of baryons and cold dark matter, massive neutrinos, photons and massless neutrinos, respectively. All perturbative quantities are computed in the N-body gauge, and the linear density field accounting for relativistic corrections is given by:
\begin{align}
    \delta \rho_{\mathrm{GR}} = \delta \rho_{\gamma}^{\mathrm{Nb}} + \delta \rho_{\rm ur} + \delta \rho_{\nu}^{\mathrm{Nb}} + \delta \rho_{\mathrm{DE}}^{\mathrm{Nb}}  + \delta \rho_{\mathrm{metric}}^{\rm Nb},
\end{align}
where the quantity $\delta \rho_{\mathrm{metric}}^{\rm Nb}$ is given by Equation (2.9) of~\cite{us2}. In this work, besides investigating the validity of COLA with respect to emulators already available in the literature, we will also analyze the effect introduced by modified gravity in the matter power spectrum, as well as its dependence on the cosmological parameters in these beyond-$\Lambda$CDM theories. To this end, we have also modified the same COLA code to consistently add purely relativistic effects from the dynamical scalar field, as well as the short scale physics introduced by it. In order to do so we further rewrite Equation (\ref{eq:COLA_impl}) as:
\begin{equation}\label{eq:COLA_impl_Geff}
      k^2 \Phi^{\rm GR} = 
     4 \pi G_{\rm eff} a^{2} \rho_{\rm m} \left[
      f_{\rm cb} \left(\delta_{\rm cb}^{(1)} + \delta_{\rm cb}^{(2)} \right)  \right] + 
      4 \pi G_{\rm N} a^{2} \rho_{\rm m}  f_{\rm GR} \delta_{\rm GR, \ rel}^{(1)},
\end{equation}
where now
\begin{align}\label{eq:d_GR_rel}
        \delta_{\rm GR, \ rel.} = \frac{\delta \rho_{\rm GR, \ rel.}}{\rho_{\rm GR}}
\end{align}
is the linear relativistic density field, defined in Equation (3.7) of~\cite{us2}, without the terms sourced by matter density perturbations in the dark energy fluctuations, which are captured by the $G_{\rm eff}$ function in Equation (\ref{eq:COLA_impl_Geff}). The geodesic equation of dark matter particles being solved is then:
\begin{equation}
    \ddot{\mathbf{x}} + 2 H \dot{\mathbf{x}} = -\nabla \Phi^{\rm GR},
\end{equation}
where for GR simulations $\Phi^{\rm GR}$ is given by Equation (\ref{eq:COLA_impl}), and for modified gravity ones $\Phi^{\rm GR}$ is given in Equation (\ref{eq:COLA_impl_Geff}).
\begin{table}[h]
    \begin{center} 
        \begin{tabular}{lc} 
            \toprule
            Parameter & Value \\
            \midrule
            Volume (Mpc$^{3}$/$h^{3}$) & $1024^{3}$  \\
            Number of particles & $1024^{3}$ \\
            Number of PM grids & $2048^3$ \\
            Initial redshift & $19$ \\
            \bottomrule						
        \end{tabular}
    \end{center}
    \caption{COLA simulation specifications.}
    \label{table:COLA_sims} 
\end{table}
Our COLA simulations have all the same specifications, shown in Table~\ref{table:COLA_sims}. Other specifications of COLA simulations used in this paper are show in Appendix~\ref{sec:AppA}. 

\subsection{Emulators}

In sight of Stage-IV LSS surveys, the efforts have been made towards producing fast and accurate theoretical predictions of summary statistics by means of emulators. Emulation methods interpolate the results of cosmological simulations in a broad range of models and cosmological parameters using machine learning techniques~\cite{agarwal2012,habib2007}. Among the various emulators produced so far, the Bacco emulator and EE2 are setting the standards in terms of the accuracy and parameters space coverage.
\begin{table}[h]
    \begin{center} 
        \begin{tabular}{lcc} 
            \toprule
            Parameter & Min. & Max. \\
            \midrule
            $h$  & $0.6$  & $0.8$ \\
            $\Omega_{\rm b}$  & $0.04$  & $0.06$ \\
            $\Omega_{\rm cdm + b}$  & $0.23$  & $0.40$  \\
            $n_{\rm s}$  & $0.92$  & $1.01$ \\
            $\sigma_{8}$  & $0.73$ & $0.9$  \\
            $\sum_{\nu} m_{\nu}$  & $0.0$ eV & $0.4$ eV \\
            $w_{0}$  & $-1.15$  & $-0.85$  \\
            $w_{a}$  & $-0.3$  & $0.3$ \\
            \bottomrule						
        \end{tabular}
    \end{center}
    \caption{Bacco training set cosmological parameter space.}
    \label{table:bacco_params} 
\end{table}
The Bacco emulator takes advantage of Principal Components Analysis (PCA) to reduce the dimensionality of the interpolation problem and applies Gaussian Process Regression to emulate each of the dimensions selected in the PCA.  It has been trained on a set of $16000$ power spectra spanning the parameter space schematised in Table~\ref{table:bacco_params}. These power spectra have been obtained using the Cosmology Rescaling algorithm~\cite{angulo2009} on a small suite of only $6$ \lcdm \ simulations obtained with L-Gadget3~\cite{gadget2,angulo2012}. The Cosmology Rescaling algorithm enables a much faster production of the training set at the expenses of a modest loss of accuracy. It has been proven to be $1\%$ accurate in \lcdm \, and $3\%$ accurate for dynamical dark energy ($w_{0}-w_{a})$ and massive neutrinos implementations. The emulator intrinsic accuracy is $\sim 2 \%$ up to scales of $5$ $h$/Mpc, so the overall accuracy is $\sim 2 \%$ in $\Lambda$CDM and $\sim 3 \%$ in the DE and massive neutrino cases. Similarly to the Bacco emulator, EE2 performs dimensionality reduction using PCA, but, then relies on a Polynomial Chaos Expansion to emulate the resulting components. The power spectra are measured at $100$ time-steps between $z=10$ and $z=0$ in a suite of $108$ pair-fixed simulations~\cite{pair_fixed} performed with PKDGRAV3~\cite{pkdgrav}. The cosmological parameters space spanned of EE2 is illustrated in Table~\ref{table:EE2_params}.%

While both emulators share some similarities in their emulation techniques, the treatment of massive neutrinos in their respective simulations is different. In the PKDGRAV3 simulations, dark energy and massive neutrinos are introduced in the same way as we have implemented in our COLA code. That is, they are introduced inside the general relativistic source term, $\delta \rho_{\rm GR}^{\rm Nb}$, as a linear density field on a Particle-Mesh (PM) grid. For the L-Gadget3 simulations of the Bacco project, they have used a cosmology rescaling algorithm~\cite{zennaro2019} to mimic the effects of massive neutrinos. This procedure is found to be $1\%$ accurate up to scales of $2$ $h$/Mpc~\cite{zennaro2019}, and $3\%$ accurate up to scales of $5$ $h$/Mpc~\cite{bacco}. However, in~\cite{bird2018}, L-Gadget3 simulations with massive neutrinos were performed using a hybrid approach, where they split massive neutrinos into ``fast'' and ``slow'' components. This allows one to combine the linear treatment of neutrinos on a PM grid, with the more accurate description of massive neutrinos as low-mass collisionless particles with large thermal velocities following CDM trajectories, which is able to incorporate in the simulations back-reaction effects of neutrinos that reduce the suppression introduced by them~\cite{bird2012}. This hybrid approach, however, agrees at below the $0.1\%$ level in the matter power spectrum with the PM method, with the only major difference between the different implementations seen in the massive neutrinos power spectrum. Therefore, for the purposes of this work, all three methods used here, EE2, Bacco and the COLA implementation are well in agreement with each other with respect to massive neutrinos.

The non-linear boost factors for the training and test sets are computed by taking the ratio of the simulations power spectra with the linear power spectra from \CLASS~\cite{class1}. The EE2 provides $\sim 1 \%$ accuracy up to $k=10 \ h/{\rm Mpc}$ in the ellipsoid centered on the reference cosmology and extending to the edges of the interpolation range.

\begin{table}[h]
    \begin{center} 
        \begin{tabular}{lccc} 
            \toprule
            Parameter & Min. & Max. & Center \\
            \midrule
            $h$  & $0.61$  & $0.73$  & $0.67$ \\
            $\Omega_{\rm b}$  & $0.04$  & $0.06$  & $0.05$ \\
            $\Omega_{\rm m}$  & $0.24$  & $0.40$  & $0.32$ \\
            $n_{\rm s}$  & $0.92$  & $1.0$  & $0.96$ \\
            $A_{\rm s}$  & $1.7\times 10^{-9}$  & $2.5\times 10^{-9}$  & $2.1\times 10^{-9}$ \\
            $\sum_{\nu} m_{\nu}$ & $0.0$ eV & $0.15$ eV & $0.075$ eV \\
            $w_{0}$  & $-1.3$  & $-0.7$  & $-1.0$ \\
            $w_{a}$  & $-0.7$  & $0.7$  & $0.0$ \\
            \bottomrule						
        \end{tabular}
    \end{center}
    \caption{EE2 parameters.}
    \label{table:EE2_params} 
\end{table}

In our work all EE2 non-linear matter power spectra shown, were computed from the linear matter power spectrum in the N-body gauge using our own version of the Einstein-Boltzmann solver \hiclass, and then multiplied by the boost factor coming from EE2~\footnote{Throughout this work we considered only the cold dark matter plus baryons power spectrum. While the Bacco emulator provides two distinct boost factors, one for the total matter and one for cold matter (cdm+b) power sepctrum, the EE2 only provides one boost factor. We have checked that the difference between the two boost factors from the Bacco emulator is negligible and we applied the EE2 boost factor to the linear cmb+b power spectrum to obtain the non-linear power spectrum.}. As EE2 has been trained using relativistic simulations with N-body gauge linear transfer functions, our choice is therefore equivalent to theirs, and for this reason in the Figures shown in this work the agreement at large scales between COLA and EE2 is well below $0.1\%$ where linear theory is valid. For the Bacco non-linear power spectra, however, we have used their own non-linear matter power spectrum prediction, that is, the linear matter power spectrum is computed inside the emulator, and the non-linear power spectra is just the Bacco boost factor multiplied by their linear prediction. This introduces a slight deviation between COLA and EE2 with Bacco at linear scales. These disagreements are expected since Bacco is not trained with relativistic simulations, nor uses the N-body gauge to compute its linear spectra. In the main text, we show comparisons with EE2 except for massive neutrinos. In Appendix~\ref{sec:AppB}, we will show comparisons with Bacco. 

\subsection{Boost and response function}
The focus of this work is to study the impact of the non-linear prescriptions by comparing different combinations of the cosmological parameters with different models of modified gravity theory and GR. Since COLA is an approximate fast method, it is not capable of predicting the non-linear power spectrum at large wave-numbers accurately. However, as we will show, COLA is capable of describing the response of the matter power spectrum with respect to the change of cosmological parameters up to $k \sim 1 \ h$/Mpc as long as the change of the matter power spectrum is not too large. Note that COLA's accuracy depends on a number of settings, such as the number of time steps and the number of grids for the PM part, and it is always a trade-off between accuracy and speed. In Appendix~\ref{sec:AppA}, we detail the specifications of COLA used in this paper. We emphasize that all the comparisons and results using our COLA simulations shown in this work can be further improved by changing these specifications at the cost of speed. In Appendix~\ref{sec:AppC}, we performed convergence tests by running high-resolution PM simulations and confirmed the robustness of our results for the response function.

Therefore, we will compare the ratio between the linear and non-linear matter power spectrum in different cosmologies with respect to a pre-defined reference cosmology, which in our case will be $\Lambda$CDM with the cosmological parameters shown in Table~\ref{table:ref_paarms}.
\begin{table}[h]
    \begin{center} 
        \begin{tabular}{lc} 
            \toprule
            Parameter & Value \\
            \midrule
            $h$  & $0.67$ \\
            $\Omega_{\rm b}$ & $0.049$ \\
            $\Omega_{\rm m}$ & $0.319$ \\
            $n_{\rm s}$ & $0.96$ \\
            $A_{\rm s}$ & $2.1 \times 10^{-9}$ \\
            $\sum m_{\nu}$ & $0.058$ eV \\
            \bottomrule						
        \end{tabular}
    \end{center}
    \caption{Reference parameters.}
    \label{table:ref_paarms} 
\end{table}
We define the linear and non-linear response functions as
\begin{align}\label{eq:R_functions}
    \Rlin(k,z) = \frac{\Plin^{\rm case}(k,z)}{\Plinref(k,z)}, \ \ \ \Rnon(k,z) = \frac{\Pnon^{\rm case}(k,z)}{\Pnonref(k,z)} 
\end{align}
respectively, where the superscript ``case" refers to a given case cosmology being investigated, and the superscript ``ref" always refers to predictions of GR with parameters from Table~\ref{table:ref_paarms}. We will also define the non-linear boost as the function that maps the linear matter power spectrum to the non-linear one
\begin{equation}
    \Pnon^{\rm case/ref}(k,z) = B^{\rm case/ref}(k,z)\times \Plin^{\rm case/ref}(k,z),
\end{equation}
and then we can get the non-linear boost in a different cosmology from the reference boost and the ratio of the response functions:
\begin{equation}\label{eq:B_function}
    B^{\rm case}(k,z) = \Bref(k,z)\times \frac{\Rnon^{\rm case}(k,z)}{\Rlin^{\rm case}(k,z)}.
\end{equation}
In the following sections, we will check the validity of using COLA to compute $\Rnon^{\rm case}(k,z)/\Rlin^{\rm case}(k,z)$ against emulators in \lcdm \ and then compute them in modified gravity models. 

\section{\lcdm \ Analyses}\label{sec:LCDM}

\subsection{Variation of cosmology parameters}\label{sec:LCDM_var_par}

In this section we compare COLA simulations with massless neutrinos with EE2 in terms of the response function defined by Equation (\ref{eq:R_functions}), by varying cosmological parameters one at a time. Throughout this work we fixed the dark energy equation of state to that of a cosmological constant, i.e., $w_{0}=-1$ and $w_{a}=0$, as well as the Hubble constant and baryon energy density to their reference values, $h=0.67$ and $\Omega_{\rm b}=0.049$.
In this section, the reference cosmology is defined by Table~\ref{table:ref_paarms} but with $\sum_{\nu} m_{\nu}=0$. 

After making these choices we are left with only three cosmological parameters, $\Omega_{\rm m}$, $n_{\rm s}$ and $A_{\rm s}$ 
, which when varied independently alongside the fixed choices of parameters, impact differently the matter power spectrum.
That is, increasing (reducing) the value of the amplitude of the primordial scalar perturbations, $A_{\rm s}$, leads to a re-scaling of the matter power spectrum amplitude up (down), while the variation of the spectral index, $n_{\rm s}$, enhances or suppresses power at small scales. Augmenting the total amount of matter in the Universe, while keeping the baryon densities fixed, leads to increasing the value of the dark matter density. This imprints the matter power spectrum by first changing the scale of equality between matter and radiation era, $k_{\rm eq}$, and by tilting the spectrum at small scales, i.e., if we have a bigger $\Omega_{\rm cdm}$ we will have steeper gravitational potentials, leading to more matter clustering at small scales, while a smaller value for $\Omega_{\rm cdm}$ gives you the opposite. 
\begin{figure}[h] 
\centering
\includegraphics[width=1\textwidth]{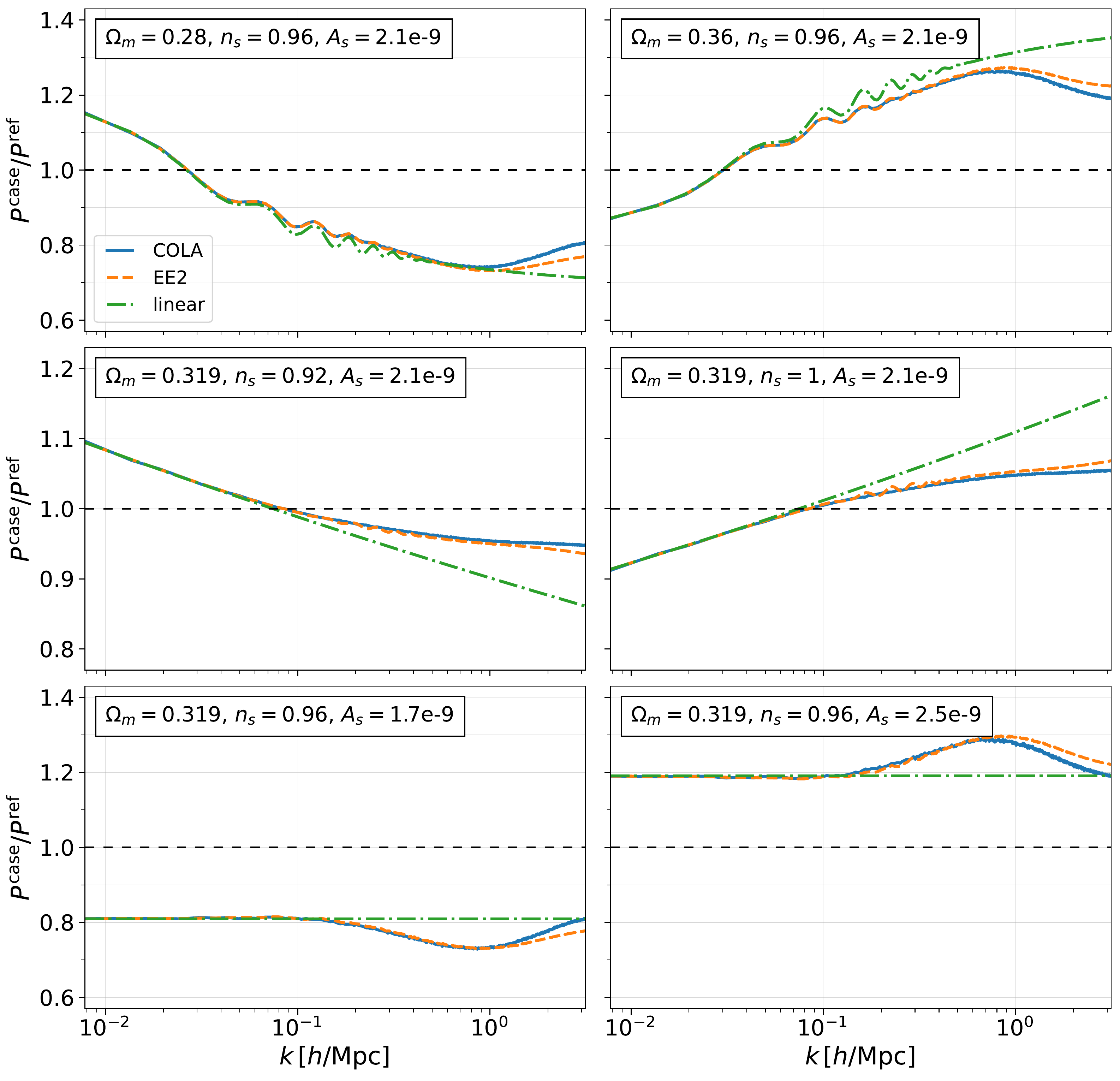}
\caption{Non-linear and linear response functions for each variation of the cosmological parameters with respect to the reference cosmology. Blue solid lines are computed from COLA simulations, orange dashed lines are obtained from EE2, and dash-dotted green lines are the linear predictions computed using \hiclass. 
}
\label{fig:cosmo_response_COLA_EE2_lin}
\end{figure}
To get a better perspective of these features, Figure~\ref{fig:cosmo_response_COLA_EE2_lin} shows the linear response and non-linear response from COLA simulations and EE2 for massless neutrinos at $z=0$. 
At large scales (small $k$ values), all curves agree with each other, while at higher $k$ values we see non-linear corrections in the solid blue and dashed orange curves.

In our COLA simulations, cosmological parameters were varied in the range shown in Table~\ref{table:varied_params}. However, in this section, we present the results and comparisons only for the cases which we will refer as ``large", that is, the minimum and maximum values shown in the same Table. 

\begin{table}[h]
    \begin{center} 
        \begin{tabular}{lcc} 
            \toprule
            Parameter & Min. & Max. \\
            \midrule
            $\Omega_{\rm m}$  & $0.28$  & $0.36$ \\
            $n_{\rm s}$  & $0.92$  & $1.0$ \\
            $A_{\rm s}$  & $1.7\times 10^{-9}$  & $2.5\times 10^{-9}$ \\
            \bottomrule
        \end{tabular}
    \end{center}
    \caption{``Large" variation of parameters.}
    \label{table:varied_params} 
\end{table}

\begin{figure}[h]
\centering
\includegraphics[width=1\textwidth]{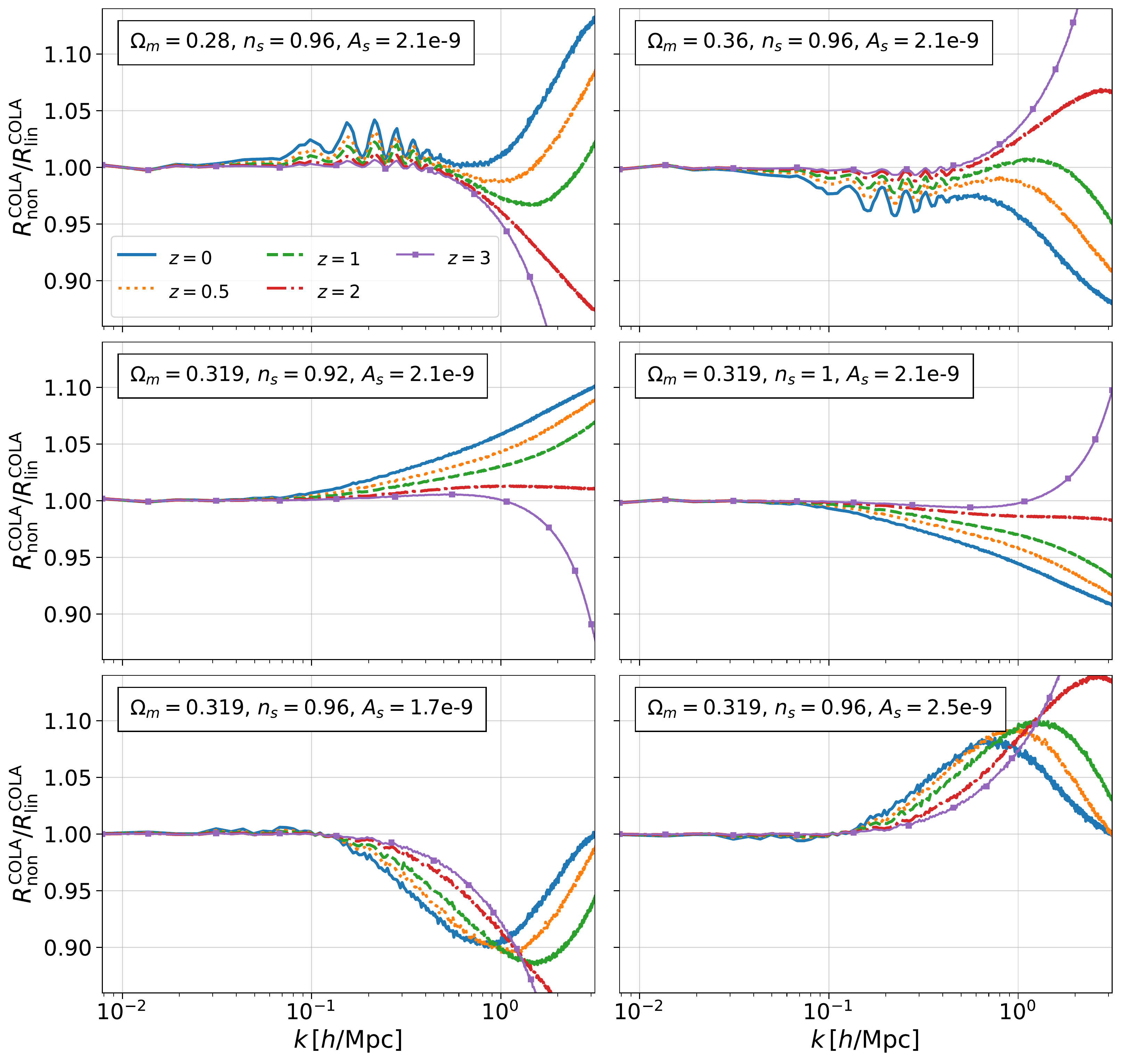}
\caption{Ratio between the non-linear and linear response functions computed from COLA simulations. We show this quantity for five different redshifts: $z=0$ (solid blue), $z=0.5$ (dotted orange), $z=1$ (dashed green), $z=2$ (dash-dotted red) and $z=3$ (solid-squared purple). We use the same convention in all other figures when we show the results at these five redshifts.}
\label{fig:mnu0_COLA_Rnon_Rlin}
\end{figure}

The difference between linear and non-linear predictions for $P^{\rm case}/P^{\rm ref}$ in  Figure~\ref{fig:cosmo_response_COLA_EE2_lin} is characterised by $R_{\rm non}/R_{\rm lin}$, which needs to be computed by simulations. 
Figure~\ref{fig:mnu0_COLA_Rnon_Rlin} show the predictions for this function by COLA at different redshifts. In our COLA implementation, the linear prediction from COLA and the one from \hiclass \ have a $0.1\%$ agreement with each other. For the change of $\Omega_{\rm m}$, we see oscillations on quasi non-linear scales, which describe the smoothing of BAO oscillations in $P^{\rm case}/P^{\rm ref}$ by non-linearity. For the change of $n_{\rm s}$ and $A_{\rm s}$, the non-linearity gives a scale-dependent enhancement or suppression at large $k$.  

\begin{figure}[h] 
\centering
\includegraphics[width=1\textwidth]{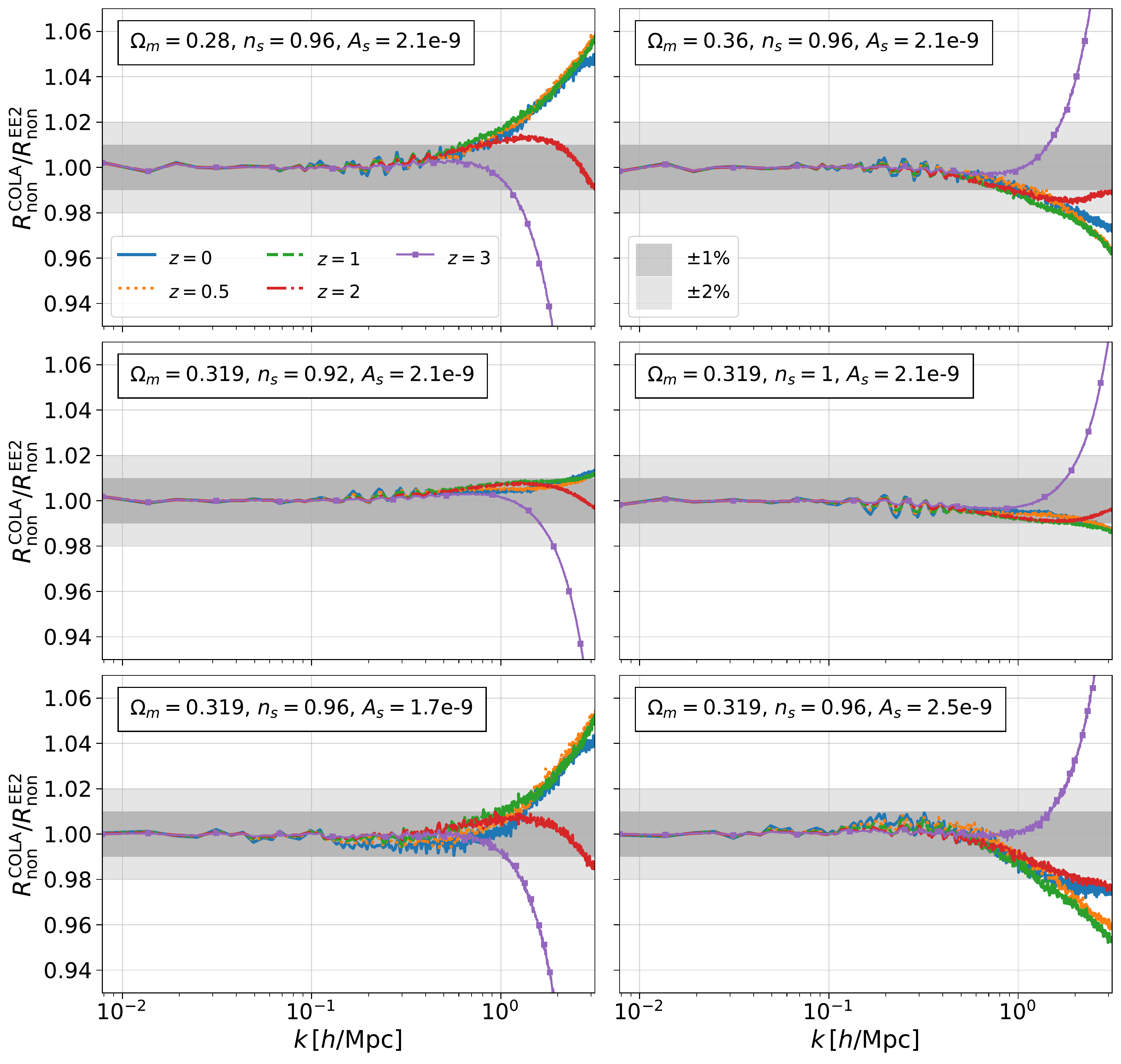} 
\caption{Ratio between the non-linear response function computed using COLA and the EE2 for the massless neutrinos case.} 
\label{fig:mnu0_COLA_EE2_ratio_Rs}
\end{figure}

To investigate how well COLA fairs with EE2 in predicting $R_{\rm non}/R_{\rm lin}$ , in Figure~\ref{fig:mnu0_COLA_EE2_ratio_Rs} we plot the ratio between the non-linear response from COLA with respect to that of EE2 for the same massless neutrinos case. We can see that we get $2\%$ agreements up to $k \sim 1 \ h$/Mpc when varying $\Omega_{\rm m}$. When we vary $n_{\rm s}$ we get $1\%$ agreements, and for $A_{\rm s}$ we obtain $2\%$ agreements at higher redshifts, while at $z \leq 1$, they become $1\%$ up to $k \sim 1 \ h$/Mpc.
In Appendix \ref{sec:AppA}, we show a comparison between EE2 and Bacco. Note that Bacco does not cover the largest $\Omega_{\rm m}$ and $A_{\rm s}$ used in this analysis. At $k < 1 \ h$/Mpc, the agreement between EE2 and Bacco is comparable to that between EE2 and COLA although the agreement is much better at $k > 1 \ h$/Mpc as expected. 

In the above studies, we consider the cases where the matter power spectrum changes up to $30 \%$ compared with the reference cosmology as shown in Figure.~\ref{fig:cosmo_response_COLA_EE2_lin}. 
The future surveys have the ability to constrain the power spectrum at $1\%$ level. In Appendix~\ref{sec:AppB} we show the results for small variations of the cosmological parameters, 
where there are $1 \%$ variations in the matter power spectrum. In this case, COLA is able to reproduce the same results for $R_{\rm non}/R_{\rm lin}$ as EE2 and Bacco with excellent precision ($< 0.1\%$) up to $k \sim 1 \ h$/Mpc. 

This shows that COLA can predict the boost factor defined in Equation (\ref{eq:B_function}) up to $k \sim 1 \ h$/Mpc as long as the deviations of the matter power spectrum from the reference cosmology are not too large. This means that COLA can be used to emulate the matter power spectrum with given  $\Bref(k,z)$ in a few chosen reference cosmologies where we require full N-body simulations to obtain the non-linear boost.  

\subsection{Inclusion of massive neutrinos}\label{sec:LCDM_mass_nus}

The observation of the flavour oscillations of neutrinos confirmed that the sum of neutrino masses is not zero. While its exact value is not yet known, massive neutrinos play an important role in cosmology. The matter power spectrum is affected by the variation of the sum of neutrino masses in two different ways. Firstly massive neutrinos introduce relativistic corrections at small $k$ values, where the bigger the value of their mass, the larger the deviation from a simulation where these species are not correctly introduced~\cite{tram}. The other imprint massive neutrinos leave in the power spectrum is the suppression of the growth of structure at small scales, which makes the growth scale-dependent. At a specific scale, called the free-streaming scale, neutrinos travel freely out of the gravitational potentials generated by cold dark matter. At the linear level, the suppression present in the dark matter power spectrum is found to be proportional to the ratio between the neutrinos and total matter energy densities~\cite{halofit}, $f_{\nu} = \Omega_{\nu}/\Omega_{\rm m}$, therefore, the larger the mass the greater the suppression. Since the two effects have their intensities related to the sum of the masses of neutrinos, this highlights the importance that neutrinos have for the next generation LSS surveys.

As both emulators used in this work have been trained assuming three degenerate massive neutrinos, we ran COLA simulations accordingly for two different values of the sum of the masses, $0.058$ eV and $0.15$ eV, and we also compare both of these cases with the massless case. 

\begin{figure}[h] 
\centering
\includegraphics[width=1.\textwidth]{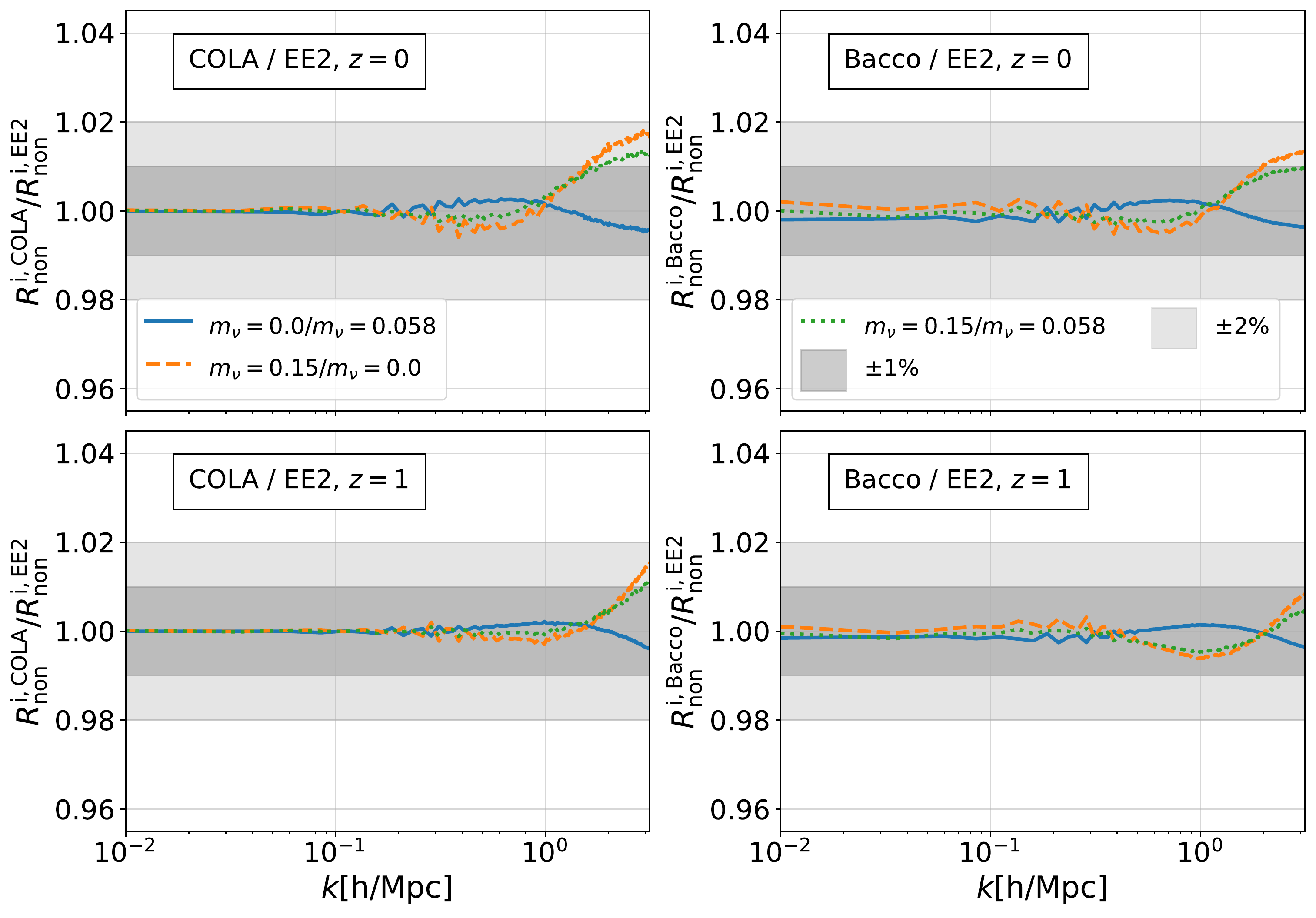}
\caption{Ratio of non-linear response function, Equation (\ref{eq:R_non_nus}), between COLA and EE2 (left column plots) and between Bacco and EE2 (right column plots) for two redshifts, $z=0$ (top row plots) and $z=1$ (bottom row plots). Solid blue lines compare the ratio $\sum_{\nu} m_{\nu}= 0.0$ eV / $\sum_{\nu} m_{\nu} = 0.058$ eV, orange dashed lines compare the ratio $\sum_{\nu} m_{\nu}= 0.15$ eV / $\sum_{\nu} m_{\nu} = 0.058$ eV, and dotted green lines compare the ratio $\sum_{\nu} m_{\nu}= 0.15$ eV / $\sum_{\nu} m_{\nu} = 0.058$ eV.}
\label{fig:mnu_COLA_EE2_bacco}
\end{figure}

With the output of our simulations we then computed the quantity
\begin{equation}\label{eq:R_non_nus}
    R^{\rm ratio, COLA}_{\rm non} = \frac{P_{\rm non}^{m_{\nu i}}}{P_{\rm non}^{m_{\nu j}}},
\end{equation}
which is the non-linear response function between two different neutrino masses, where $i,j$ refers to one of the three masses considered: $\sum_{\nu} m_{\nu}$ $=0.0$ eV, $0.058$ eV and $0.15$ eV. As we are interested in the comparison of COLA and emulators, we evaluated the same quantity for EE2 and Bacco as well. The result is plotted in Figure~\ref{fig:mnu_COLA_EE2_bacco}, which shows the ratio of the non-linear response function (\ref{eq:R_non_nus}) for each method. The curves shown in Figure~\ref{fig:mnu_COLA_EE2_bacco} show us how the non-linear suppression between different prescriptions of the matter power spectrum are in agreement with each other. From Figure~\ref{fig:mnu_COLA_EE2_bacco}, at $z=1$ we can see that COLA and EE2 have a below $0.1\%$ agreement with each other in almost the full range of scales, i.e., from small scales to beyond $1$ $h$/Mpc. While at $z=0$ the agreement slightly degrades at $k > 1$ $h$/Mpc due to the fact that EE2 is more accurate than COLA simulations at smaller scales. We find a similar agreement between Bacco and EE2 at $k < 1$ $h$/Mpc. Therefore, modulo deviations smaller than $0.5\%$ up until $k=1$ $h$/Mpc, the three methods are in excellent agreement, and the implementation of massive neutrinos in COLA does not introduce any biases when compared to the well validated emulators.

\section{Modified Gravity}\label{sec:MG}
In this section we will study how modified gravity affects the growth of structure. We will use the model-independent formalism of the EFT of DE to introduce non-linear corrections sourced by the scalar field. That is, we ran COLA simulations following Equation (\ref{eq:COLA_impl_Geff}), where the $G_{\rm eff}$ function is given by
\begin{equation}\label{eq:Geff}
    G_{\rm eff} = 1 + \frac{ c_{\mathrm{sN}}^{2} \left( 2- 2M_{*}^{2} + 2\alpha_{\mathrm{T}} \right) + \left( \alpha_{\mathrm{B}} + 2 \alpha_{\mathrm{M}} -2\alpha_{\mathrm{T}} + \alpha_{\mathrm{B}}\alpha_{\mathrm{T}} \right)^{2}}{2c_{\mathrm{sN}}^{2} M_{*}^{2}},
\end{equation}
and $\alpha_{i}$, $M_{*}^{2}$ and $c_{\rm sN}^{2}$ functions are given in the Appendix of references~\cite{us1,us2}. The common approach of modified gravity N-body codes compute $G_{\rm eff}$ by solving the scalar field fluctuation equation under the Quasi-static approximation (QSA), and then substitute the solution into the Poisson equation. In our approach, however, we are able to bypass this by splitting the dark energy perturbations into two parts: one that is purely relativistic (dynamical) and another that is sourced by matter density perturbations (QSA). In this way we are able to model the scale-dependence of the growth function present in modified gravity theories at large scales by $\delta_{\rm GR, \ rel}$ (Equation \ref{eq:d_GR_rel}) , while allowing non-linearities in matter perturbations using $G_{\rm eff}$. 

We note that there is no additional cost to include $\delta_{\rm GR, \ rel}$ (Equation \ref{eq:d_GR_rel}) in our simulations compared with simulations with massive neutrinos. We just need to use $\delta_{\rm GR, \ rel}$ including both massive neutrinos and modified gravity effects. 

Although not discussed in this work, there is one last step that is necessary to complete the description of modified gravity simulations, i.e. the introduction of screening mechanisms. In the FML implementation of COLA, screening mechanisms can be included either using screening approximations or solving the exact scalar field equation using a multi-grid solver. However, screening mechanisms are model-dependent, and, therefore, we need to first choose a modified gravity theory, and then investigate which screening mechanism will be realised in this theory. This spoils the model-independent approach we have followed in this work. For this reason, we chose not to model the shielding of the fifth-force at small scales. However, this has been well studied in the literature \cite{mgsim}, and can be easily introduced in our formalism. 

Continuing our model-independent approach, we need to select a parametrization for the time-dependent $\alpha_{i}$ functions~\cite{BS}. For the sake of generality and familiarity with previous works, we will use the following parametrization:
\begin{equation}\label{eq:alpha_param}
    \alpha_{i} = c_{i} \times a, \ \ \ \ i = \mathrm{M, \ B, \ K, \ T,}
\end{equation}
where the $c_{i}$'s are constants and $a$ is the scale factor. Throughout this work we will fix the kineticity function, $\alpha_{\rm K} = 10 \times a$, and we will focus on only two sets of Horndeski theories~\cite{horn}, one where we have $\alpha_{\rm B} \neq 0$ and $\alpha_{\rm M } = \alpha_{\rm T} = 0$, i.e., only-brading case, and the Jordan-Brans-Dicke~\cite{jbd} like case, $\bra = -\run \neq 0$ and $\ten =0$. Additionally, following the EFT of DE approach, besides fixing the $\alpha_{i}$ functions, we are still left to choose one more function that fully characterizes the evolution of dark energy models in a model-independent framework, the background evolution history, $H(a)$. For simplicity, and due to the multiple constraints one can get on the expansion of our Universe, we choose to fix it to an $\Lambda$CDM evolution with the energy fractional densities given by Table~\ref{table:ref_paarms} or Table~\ref{table:varied_params}. 
We emphasize that these functions are chosen for illustrations of our approach and any functions can be used in simulations. 

\begin{figure}[h] 
\centering
\includegraphics[width=1.\textwidth]{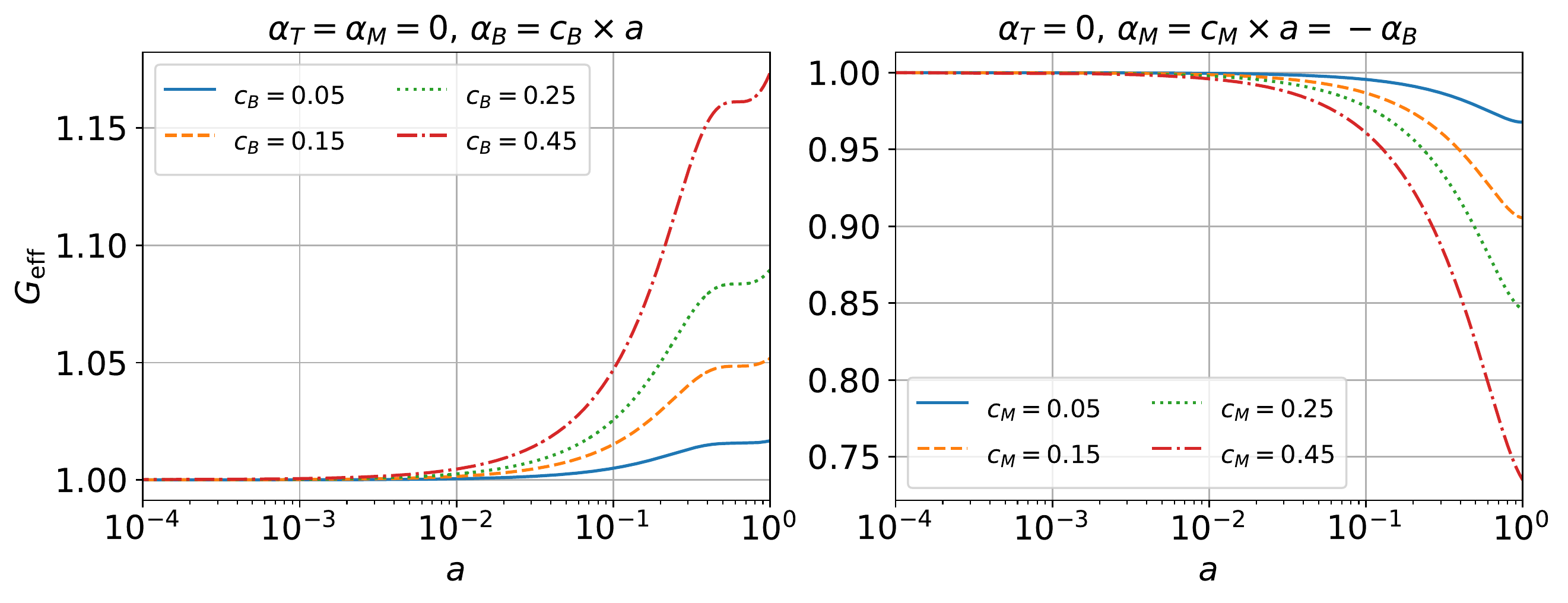} 
\caption{\textbf{Left:} Evolution of the $G_{\rm eff}$ function in the ``only-braiding" gravity model i.e. $\bra = c_{\rm B} \times a$ and $\ten=\run=0$, for $c_{\rm B}=$ $0.05$ (solid blue), $0.15$(dashed orange), $0.25$ (dotted green) and $0.45$ (dash-dotted red). 
\textbf{Right:} Evolution of the $G_{\rm eff}$ function in the ``JBD-like" gravity model, i.e. $\bra = -\run = c_{\rm M} \times a$ and $\ten=0$, for $c_{\rm M}=$ $0.05$ (solid blue), $0.15$ (dashed orange), $0.25$ (dotted green) and $0.45$ (dash-dotted red). In both cases $\kin = 10 \times a$.}
\label{fig:Geff_cB_JBD}
\end{figure}

We show in Figure~\ref{fig:Geff_cB_JBD} the evolution of $G_{\rm eff}$ for the two models for different values of the proportionality constants. As shown in~\cite{us1,us2}, the kineticity function affects the matter power spectrum only at sufficiently large scales, $k \sim 10^{-3}$ $h$/Mpc, and due to the size of our simulations, shown in Table~\ref{table:COLA_sims}, the specific value and functional form for $\kin$ is not relevant for the exposition of our results, hence we keep it fixed. As discussed in Section~\ref{sec:method}, our COLA simulations incorporate the relativistic effects of photons, neutrinos and dark energy using the N-body gauge approach, following the same approach used by EE2. In all COLA simulations in this modified gravity section we fix the sum of neutrinos masses to $0.058$ eV.

\subsection{Fixed cosmology}\label{sec:MG_fixed_cosmo}

To investigate the interplay between modified gravity (MG) and the cosmological parameters we will first compare the linear and non-linear response, $\Rlin^{\rm MG}$ and $\Rnon^{\rm MG}$, for the modified gravity cases mentioned above, while keeping the cosmological parameters fixed to their reference values shown in Table~\ref{table:ref_paarms}.

\begin{figure}[h] 
\centering
\includegraphics[width=1\textwidth]{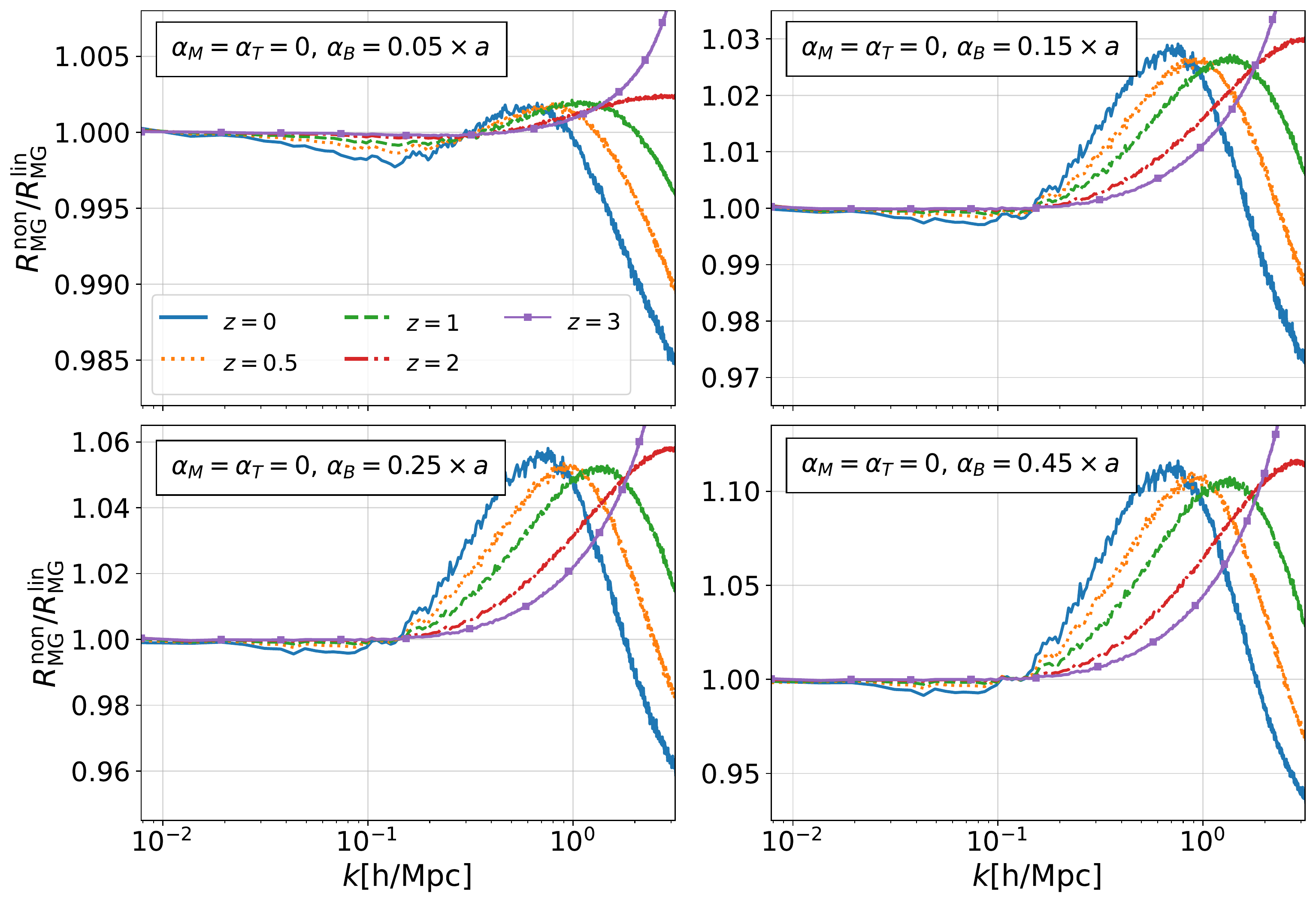} 
\caption{Ratio between non-linear and linear response functions in modified gravity (MG) ``only-braiding" models with different proportionality constants $c_{\rm B}$. 
}
\label{fig:R_non_lin_cB}
\end{figure}

\begin{figure}[h] 
\centering
\includegraphics[width=1\textwidth]{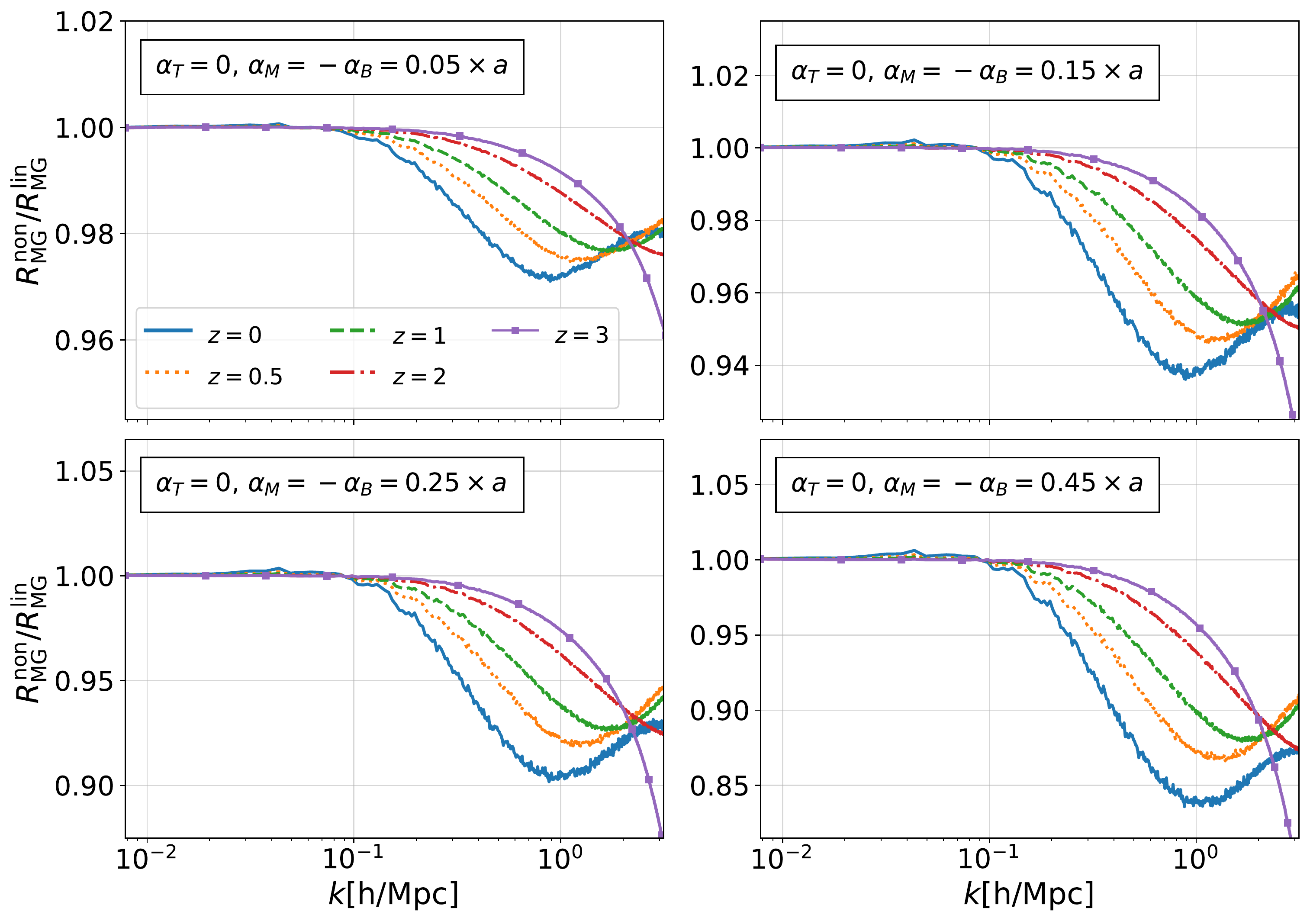} 
\caption{Ratio between non-linear and linear response functions in modified gravity ``JBD-like" models with different proportionality constants $c_{\rm M}$. 
}
\label{fig:R_non_lin_JBD}
\end{figure}

From Figures~\ref{fig:R_non_lin_cB} and \ref{fig:R_non_lin_JBD} we can see that non-linear effects are stronger for larger values of the constant of proportionality, in accordance with the plots shown in Figure~\ref{fig:Geff_cB_JBD}. More interestingly, from the top left plot of Figure~\ref{fig:R_non_lin_cB}, the lowest value of $c_{\rm B}$, we see almost no non-linear corrections in the response function up to $k \sim 1$ $h$/Mpc. In the ``only-braiding" gravity model, non-linearity introduces an enhancement, while in the ``JBD-like" model, non-linearity gives an additional suppression at large $k$. At small $k$, linear and COLA predictions are in excellent agreement due to the correct implementation of the relativistic effects in COLA simulations described in Section~\ref{sec:method}.

\begin{figure}[h] 
\centering
\includegraphics[width=1\textwidth]{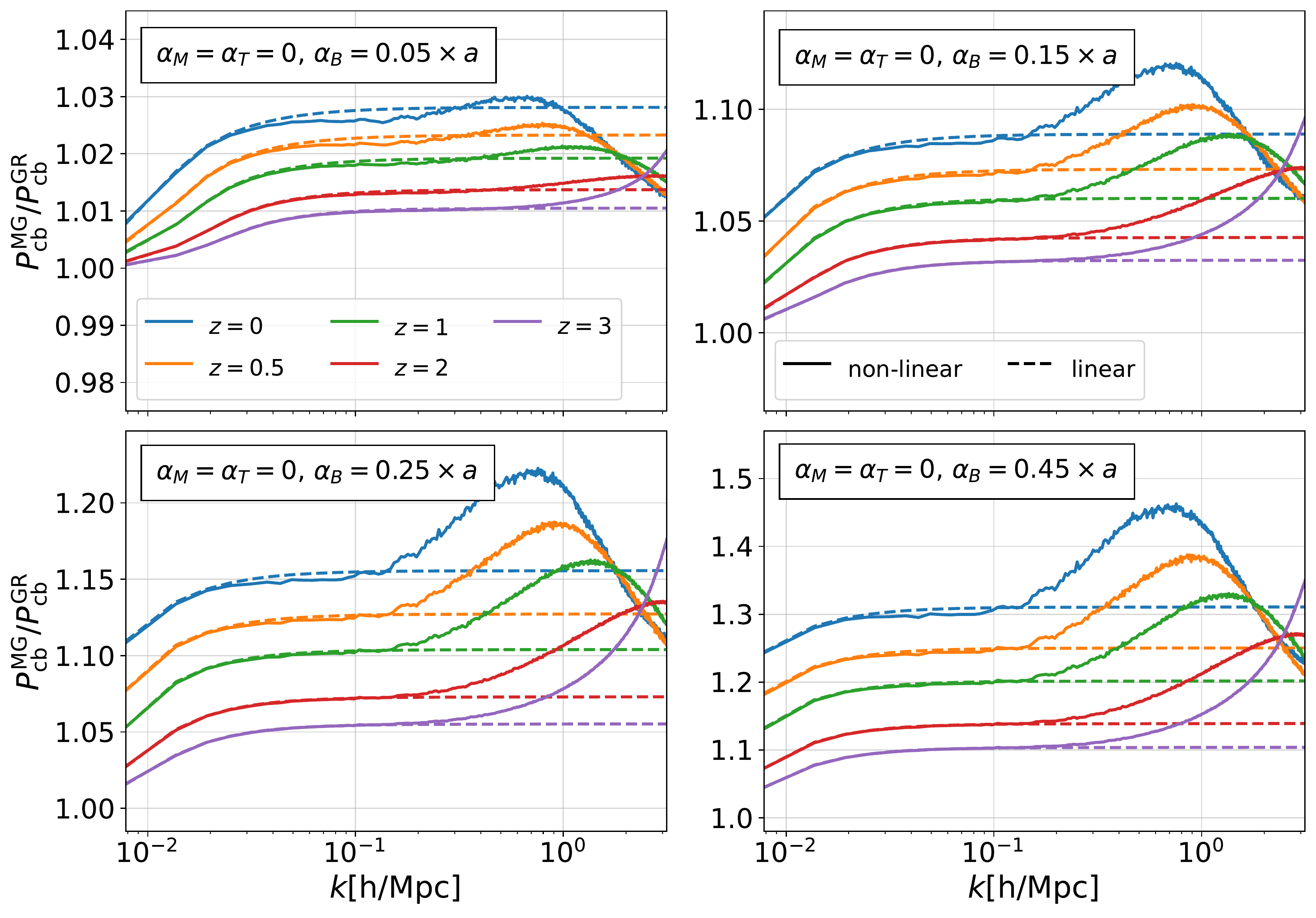} 
\caption{Ratio between modified gravity and GR power spectra in the ``only-braiding" model. 
The solid lines refer to the non-linear prediction, while the dashed ones to the linear one. 
}
\label{fig:P_cB_nl_lin}
\end{figure}

\begin{figure}[h] 
\centering
\includegraphics[width=1\textwidth]{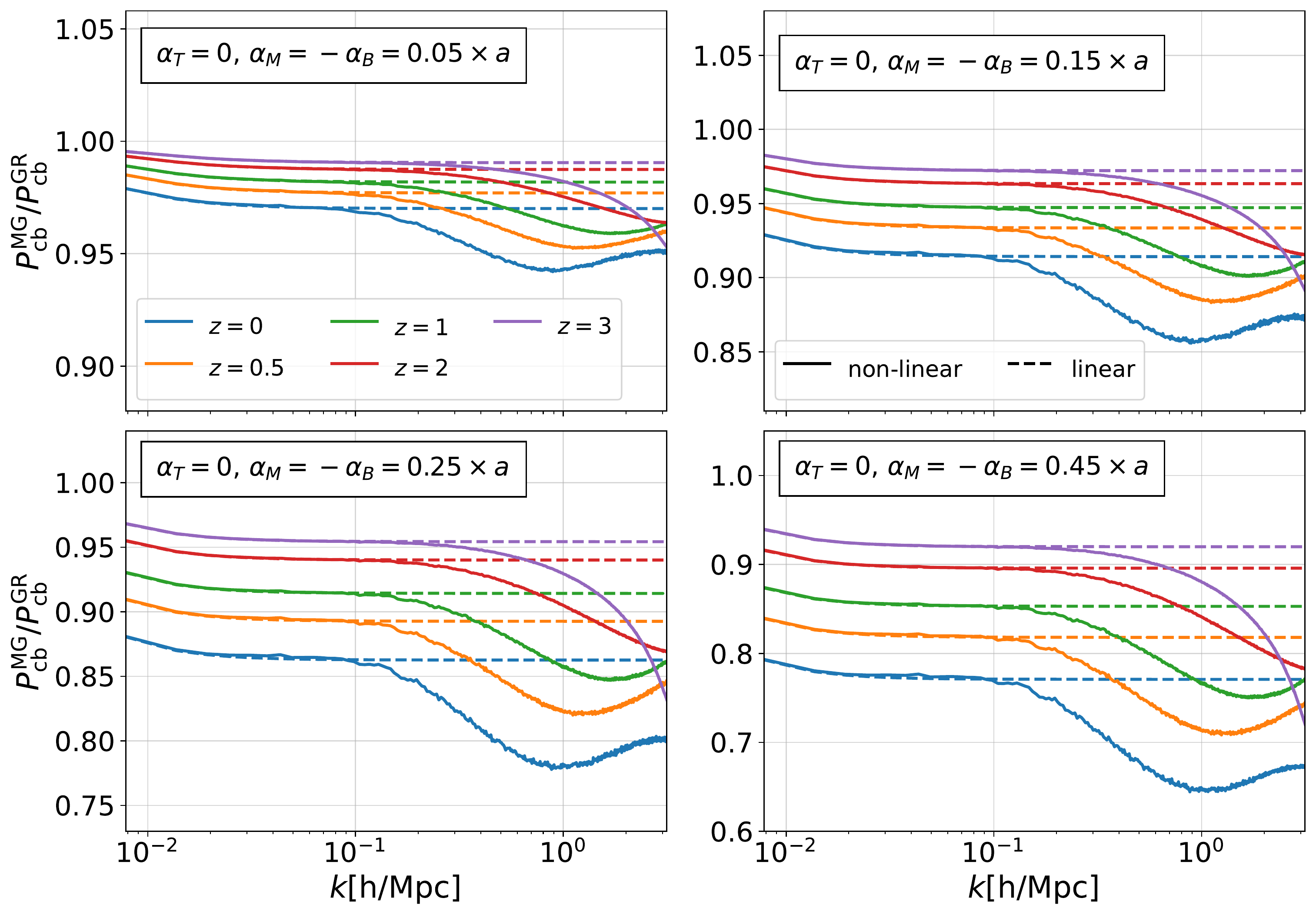} 
\caption{Ratio between modified gravity and GR power spectra in the ``JBD-like" model. 
The solid lines refer to the non-linear prediction, while the dashed ones to the linear. 
}
\label{fig:P_JBD_nl_lin}
\end{figure}

The impact of non-linear corrections in the matter power spectrum for each model discussed is shown in Figures~\ref{fig:P_cB_nl_lin} and \ref{fig:P_JBD_nl_lin}. We plot the ratio between the matter power spectrum in modified gravity with respect to the one computed using GR with the same cosmological parameters. If we use the quasi-static approximation and linear approximations, we see a constant off-set caused by $G_{\rm eff}$. 

The deviation from the constant offset at small $k$ arises from relativistic corrections that are computed by the Boltzmann code. We see that at small $k$ values, the non-linear (solid lines) and linear curves (dashed lines) have the same behavior, as expected, since in our COLA implementation we have consistently introduced relativistic corrections. 

At large wave-numbers, the deviations from the constant off-set arise from non-linear effects, which are more prominent at smaller redshifts and for larger values of the modified gravity parameters, $c_{\rm B}$ (only-braiding) and $c_{\rm M}$ (JBD) as expected as non-linearity is stronger.
We see that non-linearity gives an additional enhancement at large $k$ in the ``only-braiding" case while it gives an additional suppression in the ``JBD" model. As shown in the previous section, COLA should be able to capture these non-linear corrections accurately up to $k \sim 1 \ h$/Mpc. 

\subsection{Varying cosmology}\label{sec:MG_var_cosmo}
We now move to the next discussion where we will vary the cosmological parameters $A_{\rm s}$, $n_{\rm s}$ and $\Omega_{\rm m}$ as shown in Table~\ref{table:varied_params}. Our main goal in this section is to investigate the dependence of modified gravity effects on these parameters. Since the effect of the Horndeski scalar field seen in Figures~\ref{fig:R_non_lin_cB} and \ref{fig:R_non_lin_JBD} is very similar in different models modulo the difference between enhancing/suppressing the growth, we chose to select only one case of modified gravity, i.e. the only-braiding case with $c_{\rm B} = 0.45$. This value is of particular interest as the non-linear response with respect to the linear one is roughly $10\%$ at $z=0$, and we will refer to this value as ``cB10". Note that the effect on the matter power spectrum is around $30 \%$ on linear scales at $z=0$.  

\begin{figure}[h] 
\centering
\includegraphics[width=1\textwidth]{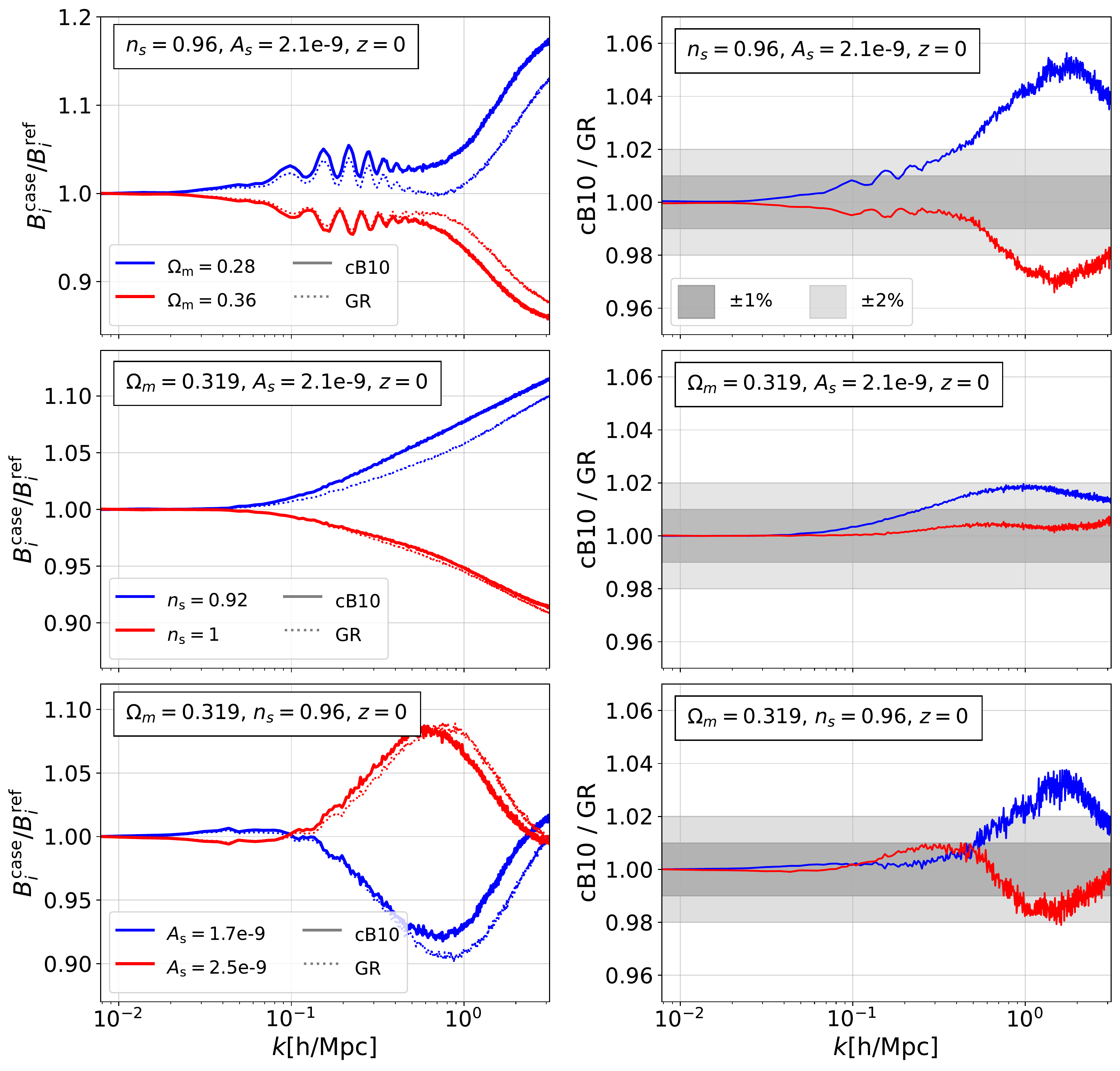} 
\caption{In the left panel, we show the ratio between the boost factor in each ``large'' cosmological parameter variation case and that in the reference cosmology. The solid lines show this quantity in the modified gravity ``only-braiding" case (cB10) while dotted ones show this in GR. In the right panel, we show the ratio of this ratio between modified gravity and GR. Top, middle and bottom row plots show variations with respect $\Omega_{\rm m}$, the spectral index $n_{\rm s}$ and the amplitude of the primordial perturbations $A_{\rm s}$, respectively.
}
\label{fig:B_case_B_ref_cb10_GR}
\end{figure}

In the left hand side of Figure~\ref{fig:B_case_B_ref_cb10_GR}, we plot $B^{\rm case}/B^{\rm ref}$ in the reference cosmology and the model with ``cB10''. The superscript ``case" refers to one of the $6$ cases of ``large'' values in Table~\ref{table:varied_params}.
The right hand side of the plot shows the ratio: 
\begin{equation}\label{eq:curvy_R0}
    \mathcal{R}_{\rm cB10, GR} = \frac{B_{\rm cB10}^{\rm case}}{B_{\rm cB10}^{\rm ref}} \times \left(\frac{B_{\rm GR}^{\rm case}}{B_{\rm GR}^{\rm ref}}\right)^{-1}
\end{equation}
We can see that all cases where the variation of the cosmological parameters decreased their values from the reference ones, the ratio $B^{\rm case}/B^{\rm ref}$ increase in the ``cB10'' model compared with the reference cosmology in GR.
Note that in these cases, the effect of changing these cosmological parameters is to decrease the amplitude of the power spectrum on small scales while the effect of the modified gravity considered here is to enhance the amplitude. The increase in the ratio $B^{\rm case}/B^{\rm ref}$ is more prominent in the cases of the parameters $\Omega_{\rm m}$ and $A_{\rm s}$, as these two are related to changes in the overall amplitude of the power spectrum, while it is milder in the case of the spectral index reaching at most an increase of $2\%$. When the cosmological parameters increase, we see an opposite effect. Overall, compared with $B^{\rm case}/B^{\rm ref}$ in the reference cosmology, the effect of modified gravity is fairly weak at $k < 1 \ h$/Mpc except for $\Omega_{\rm m}$.

The ratio $\mathcal{R}_{\rm cB10, GR}$ can be rewritten  as
\begin{equation}\label{eq:curvy_R}
    \mathcal{R}_{\rm cB10, GR} = \frac{R^{\rm case}_{\rm non,cB10}}{R^{\rm case}_{\rm lin,cB10}} 
    \times \left(    
      \frac{R^{\rm ref}_{\rm non, cB10}} {R^{\rm ref}_{\rm lin, cB10}}
    \right)^{-1}
    \end{equation}
where
\begin{align}
    &R^{\rm case}_{\rm non,cB10} = \frac{P^{\rm case}_{\rm non, cB10}}{P^{\rm ref}_{\rm non, GR}}, \ \ \ R^{\rm ref}_{\rm non, cB10} = \frac{P^{\rm ref}_{\rm non, cB10}}{P^{\rm ref}_{\rm non, GR}}, \\
    &R^{\rm case}_{\rm lin,cB10} = \frac{P^{\rm case}_{\rm lin, cB10}}{P^{\rm ref}_{\rm lin, GR}}, \ \ \ R^{\rm ref}_{\rm lin, cB10} = \frac{P^{\rm ref}_{\rm lin, cB10}}{P^{\rm ref}_{\rm lin, GR}}
\end{align}
which is line with our previous definitions in Equations (\ref{eq:R_functions}), where the reference non-linear and linear matter power spectra are always computed with cosmological parameters in Table~\ref{table:ref_paarms} \emph{and} GR as the gravity theory. Thus $\mathcal{R}_{\rm cB10, GR}$ can be interpreted in two ways: how modified gravity changes the cosmological parameter dependence of the boost factor as discussed above, and 
how cosmological parameters affect the response of the matter power spectrum to modified gravity parameters. 
As we have seen, the cosmological parameter dependence 
of $\mathcal{R}_{\rm cB10, GR}$ is fairly weak at $k < 1 \ h$/Mpc except for $\Omega_{\rm m}$. This property is useful when creating emulators for the response functions for modified gravity parameters. This is consistent with what was found in $f(R)$ gravity in~\cite{hans_emu}. 

It is worth making some remarks about how general our results are when contrasted to the great freedom we have at choosing a parametrization for the $\alpha_{i}$ functions. For late-time dark energy models, the choice of parametrization is such that as the Universe evolves, the impact dark energy has on the matter power spectrum also increases. In this work, we have chosen to make these property functions proportional to the scale factor. This was done in order to better highlight the new features introduced by the scalar field, since making the $\alpha_{i}$ functions depending linearly on the scale factor allows them to have an impact at an earlier period of time than the case where they were proportional to the background fractional dark energy density, $\propto \Omega_{\rm DE}$, another common choice in the literature. This parametrization forces the scalar field density perturbations to only be relevant at redshifts during dark energy domination era, $z \lesssim 0.7$. 
However, the same discussions and conclusions in the present work would still be valid. 

\section{Conclusion}\label{sec:concl}

In this work, we have validated COLA simulations with two existing emulators in the literature, the EE2 and the Bacco emulator. In order to do so, we have made use of the non-linear response function, defined in Equation (\ref{eq:R_functions}), where the reference cosmology is given in Table~\ref{table:ref_paarms}. Each case cosmology corresponded to the variation of one of the following cosmological parameters: the total matter energy density, spectral index and amplitude of the primordial fluctuations, within the range detailed in Table~\ref{table:varied_params}. All COLA simulations followed the same specifications presented in Table~\ref{table:COLA_sims} and Appendix~\ref{sec:AppA}. The performance of COLA can be improved further at the cost of speed. 

In Section~\ref{sec:LCDM_var_par} we have analysed how the ratio between the non-linear and linear response functions, computed using COLA, is affected by varying the cosmological parameters to large deviation values, i.e. the boundaries of the ranges shown in Table~\ref{table:varied_params} as displayed in Figure~\ref{fig:mnu0_COLA_Rnon_Rlin}. To investigate the agreement between COLA and EE2, in Figure~\ref{fig:mnu0_COLA_EE2_ratio_Rs} we showed the ratio between the non-linear response function computed using each method, and we concluded that both predictions agree at the $2\%$ level up until $k=1 \ h/$Mpc. For small variations of the cosmological parameters (0.5$\%$ variations), the agreement is even better, being well below the $0.1\%$ level as shown in Appendix~\ref{sec:AppB}, which also includes the validation with the Bacco emulator.

To test the implementation of massive neutrinos in COLA, in Section~\ref{sec:LCDM_mass_nus}, we computed the non-linear suppression computed by each non-linear prescription, i.e. Equation (\ref{eq:R_non_nus}), at $z=0$ and $z=1$. We used three different values of the sum of the neutrino masses, $0.0$ eV, $0.058$ eV and $0.15$ eV, in our COLA simulations. In Figure~\ref{fig:mnu_COLA_EE2_bacco}, we plotted the ratio of the non-linear suppression between the different cases of neutrino masses evaluated by each method, COLA, EE2 and Bacco. We found that all three were in agreement at below $0.5\%$ level at both redshifts down to $k = 1 \ h/$Mpc, showing that the treatment of massive neutrinos in COLA does not introduce any biases, and the non-linear response function computed using COLA simulations can be used to extend existing emulators in the literature, as well as to train new ones.

In Section~\ref{sec:MG} we introduced our implementation of scalar-tensor theories of gravity in COLA via the model-independent approach of the EFT of DE. In order to do so, all non-linearities arising from the scalar field fluctuations are encoded in the $G_{\rm eff}$ function, Equation (\ref{eq:Geff}), which functional form is found by assuming that non-linear corrections of modified gravity are sourced only by matter density perturbations (QSA limit). At the same time, we also implemented relativistic corrections from the scalar field perturbations via the N-body gauge approach developed in~\cite{us1,us2}. 

In Figures~\ref{fig:R_non_lin_cB} and \ref{fig:R_non_lin_JBD} we showed how non-linearities affect the response function in two different models, the only-braiding model and the JBD-like one. The main effect corresponds to a rescaling of the amplitude of the matter power spectrum by either enhancing the power at small scales in the only-braiding case, or suppressing the growth of structure at these scales in the JBD-like model as shown in  Figures~\ref{fig:P_cB_nl_lin} and \ref{fig:P_JBD_nl_lin}. These figures showed also that our COLA simulations included dynamical effects of relativistic species including the scalar field on large scales as well as non-linear clustering of dark energy driven by non-linear matter perturbations on smaller scales.

Based on this analysis, in Section~\ref{sec:MG_var_cosmo}, we have then chosen a specific case of the only-braiding gravity, ``cB10", to investigate how modified gravity impacts the variation of cosmological parameters. This was done by running new COLA simulations for large variation values of $\Omega_{\rm m}$, $A_{\rm s}$ and $n_{\rm s}$, but with the gravity theory being described by ``cB10". As showed in Figure~\ref{fig:B_case_B_ref_cb10_GR}, our results showed that the dependence of the response function with respect to the change of the modified gravity parameter on $A_{\rm s}$ and $n_{\rm s}$ is fairly weak up to $k=1 \ h/$Mpc. For $\Omega_{\rm m}$, we saw a stronger dependence, which was also found in reference~\cite{hans_emu}. 

With the results of our investigations, we conclude that COLA simulations can be used to extend emulators already available in the literature, as well as to train new ones, via the use of the response functions computed from these simulations. This method is able to push the regime of validity of COLA down to $k=1 \ h/$Mpc with accuracy below $1\%$ if the deviations from the reference cosmology are small enough. For beyond-$\Lambda$CDM cosmologies this is particularly important, as running COLA simulations are much faster than full N-body ones. Also, since the dependence on cosmological parameters is fairly mild in the response function with respect to modified gravity parameters, we can use cosmology-independent methods to include new theories of gravity into emulators, which is a very desirable feature for the upcoming LSS surveys.

In order to go beyond $k=1 \ h/$Mpc, we need to improve the PM part of the simulations. There are several methods proposed to improve the accuracy of COLA and PM simulations \cite{necola,dai_2018,lanzieri_2022}. In addition, baryonic effects become significant beyond $k=1 \ h/$Mpc~\cite{chisari_2019} and these effects need to be added to dark matter only simulations using the methods proposed by \cite{schneider,dai_2018} for example. It is still an open question whether we gain any information on cosmology and modified gravity by marginalising over parameters describing these baryonic effects beyond $k=1 \ h/$Mpc. Even for emulators using more sophisticated N-body codes to go beyond $k=1 \ h/$Mpc such as the FORGE emulator in $f(R)$ gravity \cite{forge}, our study indicates that it is easier to emulate the response function to avoid the effect of resolution issues. Finally, there is a halo model approach, ReACT~\cite{bose_2020,bose_2021}, to predict the non-linear power spectrum in modified gravity models. This method requires a calibration of the concentration-mass relation for the 1-halo term to get a precise agreement with N-body results~\cite{srinivasan_2021}. It will be interesting to compare our results with the prediction of ReACT for the models considered in this paper.

\acknowledgments
For the purpose of open access, the authors have applied a Creative Commons Attribution (CC BY) licence to any Author Accepted Manuscript version arising from this work. 
GB acknowledges support from the State Scientific and Innovation Funding Agency of Esp\'irito Santo (FAPES, Brazil), and is supported by the Alexander von Humboldt Foundation. KK is supported by the UK STFC grant ST/S000550/1 and ST/W001225/1. Numerical computations were done on the Sciama High Performance Compute (HPC) cluster which is supported by the ICG, SEPNet and the University of Portsmouth. 

\paragraph{Data availability} Supporting research data are available on reasonable request from the corresponding author.

\appendix 

\section{COLA specifications}\label{sec:AppA}

COLA is a quasi N-body code, that is, it is an approximate method that allows us to reduce the number of time-steps that usual full N-body simulations require getting faster realizations of the non-linear density field. Therefore, if we want to get more accurate descriptions of clustering on small scales, we can just increase the number of time-steps in our COLA simulations to make it closer to full N-body codes. However, there is an obvious setback to this, as we increase time-steps we increase the time our simulations take to finish. Also, the initial redshift of these simulations impacts the large $k$ behavior as well, since the higher the redshift we choose, the more time-steps between $z_{\rm ini}$ to smaller redshifts we will need. Another parameter COLA is sensitive to, is the number of cell grids in its PM algorithm. As COLA solves the Poisson equation using inverse and normal FFTs, by increasing the number of cells we will be making our mesh finer and increasing the force resolution. 

These specifications used in this work follow closely the discussion presented in~\cite{albert} where detailed analyses were performed to find the best specifications for COLA simulations. All our simulations use the box size of $1024 \ h/$Mpc with $N_p^3= 1024^3$ particles.  

In Table~\ref{table:COLA_sims} we chose to use $50$ time-steps for our COLA simulations starting at $z_{\rm ini} = 19$, with the time-steps subdivided as shown in Table~\ref{table:COLA_ts}.
\begin{table}[h]
    \begin{center} 
        \begin{tabular}{lc} 
            \toprule
            Redshift & Number of time-steps \\
            \midrule
            $19 \to 3$  & $12$ \\
            $3 \to 2$  & 5 \\
            $2 \to 1$  & $8$ \\
            $1 \to 0.5$  & $9$ \\
            $0.5 \to 0$  & $17$ \\
            \bottomrule
        \end{tabular}
    \end{center}
    \caption{Number of time-steps intervals.}
    \label{table:COLA_ts} 
\end{table}
These time steps are linearly distributed in the scale factor. As we can see from this table that the COLA method generally uses a very sparse time-stepping at higher redshifts, which can be a problem if we have a very low force resolution. In order to not have a loss of power at small scales in our simulations, we then chose to use a force mesh grid number of $N_{\rm mesh}^3 = (2 N_p)^3 =2048^3$ as it was shown that increasing this to $(3 N_p)^3$ has $\sim 1 \%$ effects on the matter power spectrum at $k < 1 \ h$/Mpc. 

\begin{figure}[h] 
\centering
\includegraphics[width=.7\textwidth]{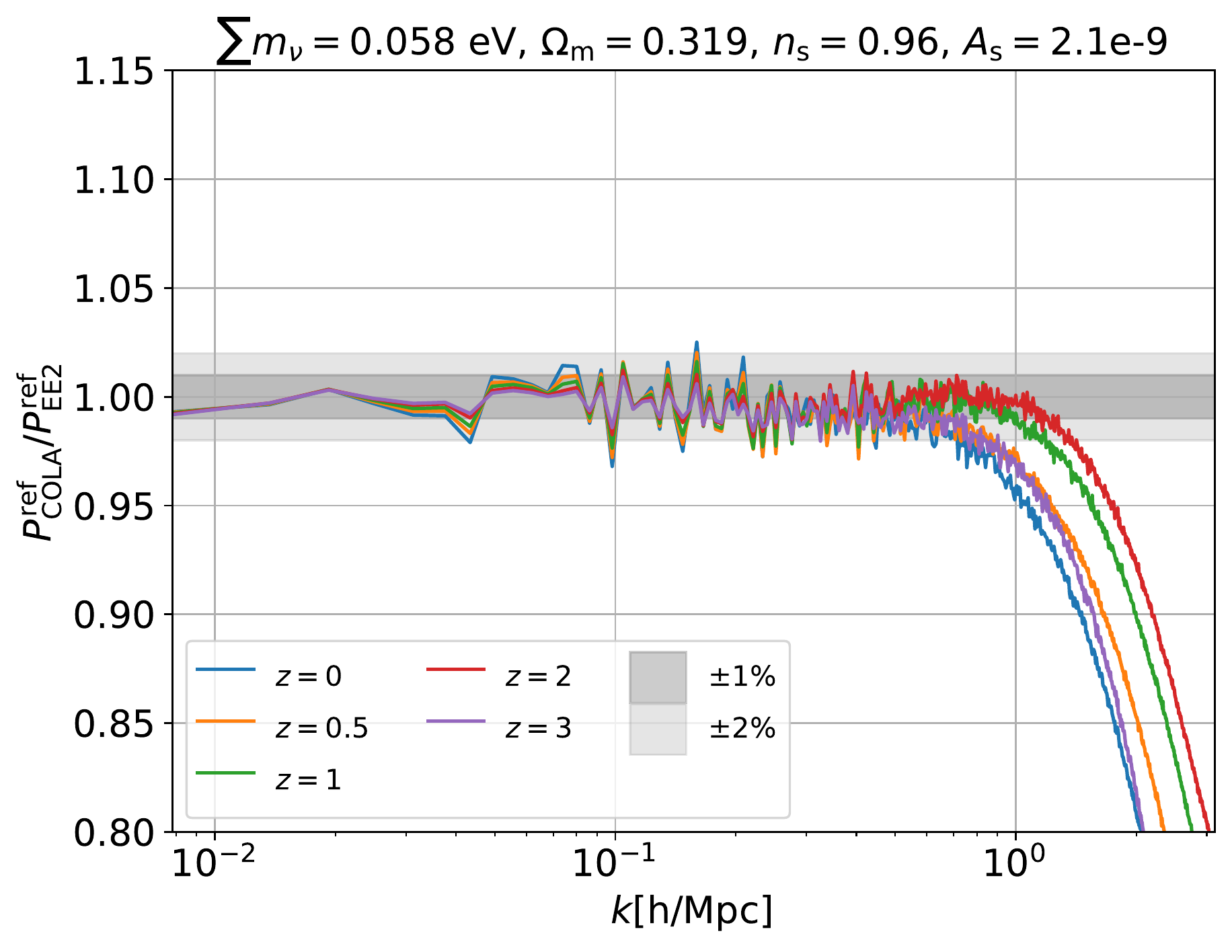} 
\caption{Ratio between the absolute non-linear dark matter power spectrum computed using COLA and EE2 for the reference cosmology.} 
\label{fig:P_nl_COLA_EE2_ref}
\end{figure}

In Figure~\ref{fig:P_nl_COLA_EE2_ref} we show the ratio between the absolute non-linear power spectrum measured from our COLA simulation using the reference values of Table~\ref{table:ref_paarms}, and the EE2 non-linear matter power spectrum. We use the paired-fixed simulations to reduce the cosmic variance. 
We see that the ratio between the two is highly oscillatory when we go to large $k$ values, due to the residual sample variance effects of our COLA simulation. 
The time steps are chosen to reproduce the matter spectrum better at $z \leq 1$. At $z \leq 3$, the non-linear power spectrum from COLA agrees with EE2 at $1 \%$ level at $k \leq 0.5 \ h$/Mpc. As shown in this paper, COLA gives a better accuracy up to higher $k$ in predicting reactions of the matter power spectrum to the change of cosmological parameters as well as massive neutrino mass. Also since the reaction is defined as a ratio of the power spectrum, using the same initial seed, we can suppress the sample variance and obtain much smoother predictions. To compute the reaction, we used fixed amplitude simulations and the same initial seed for all the simulations. 

We emphasize that all the comparisons and results using our COLA simulations shown in this work can be even further improved by increasing the temporal resolution of our simulations, i.e., increasing the total number of time-steps, the force resolution, i.e., increasing the the number of cell grids in our PM algorithm, and the particle number. In appendix we performed convergence tests 

\begin{figure}[h] 
\centering
\includegraphics[width=1.0\textwidth]{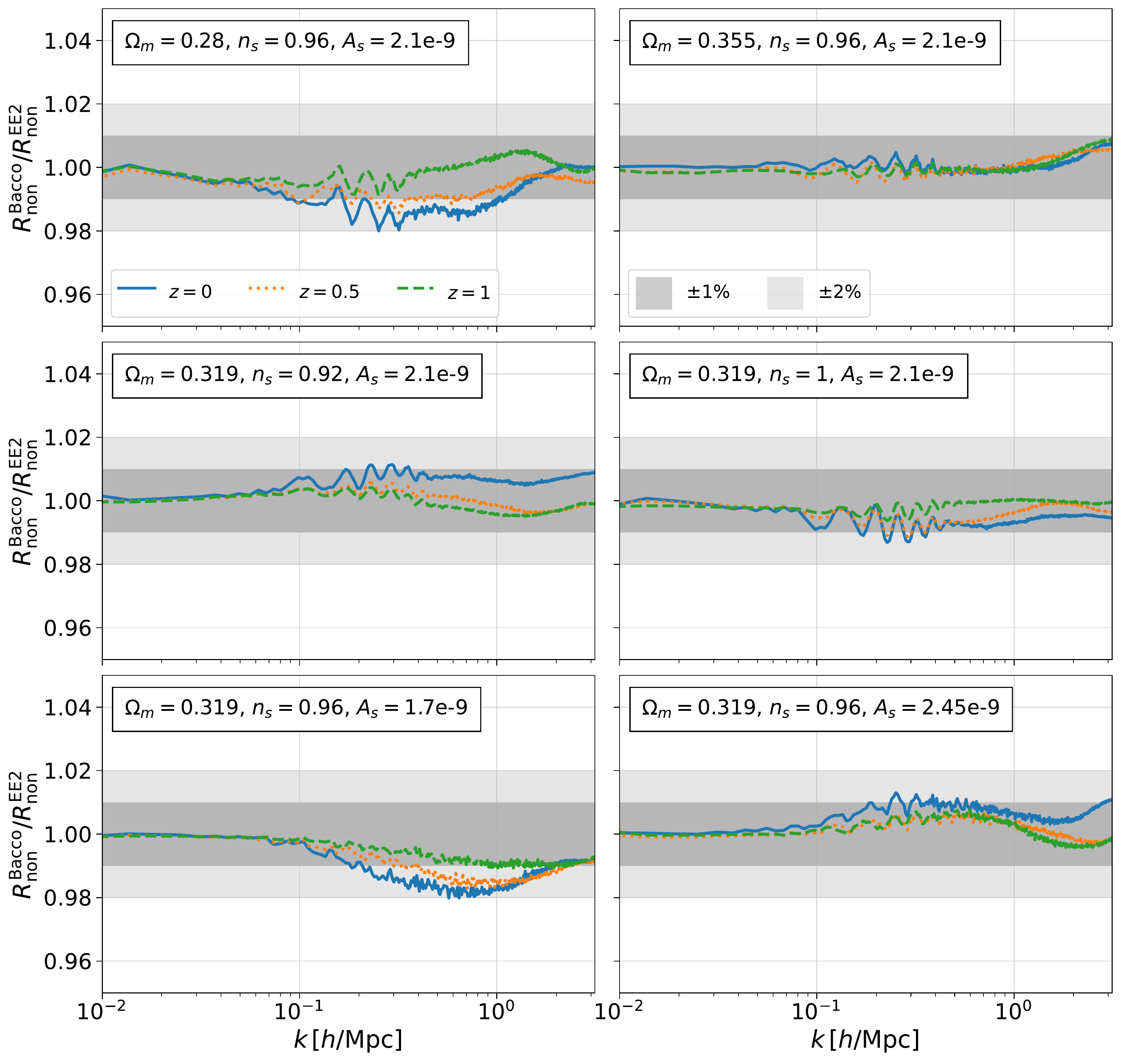} 
\caption{Ratio between the non-linear response function computed using Bacco and the EE2 for the massless neutrinos case. }
\label{fig:Rnon_BaccovsEE2_LargeVar}
\end{figure}

Finally, in the main text, we only showed comparisons between COLA and EE2 for the prediction of the reaction. Here, for completeness, we show a comparison between EE2 and Bacco in Figure~\ref{fig:Rnon_BaccovsEE2_LargeVar}. As noted in the main text, Bacco do not cover the largest values of $\Omega_{\rm m}$ and $A_{\rm s}$ used in our analysis. In the plot, we used $\Omega_{\rm m}=0.355$ and $A_{\rm s}=2.45 \times 10^{-9}$ instead. We get below $2\%$ agreements in all cases.  

\section{Small variations of cosmological parameters}\label{sec:AppB}
\begin{figure}[h] 
\centering
\includegraphics[width=1.0\textwidth]{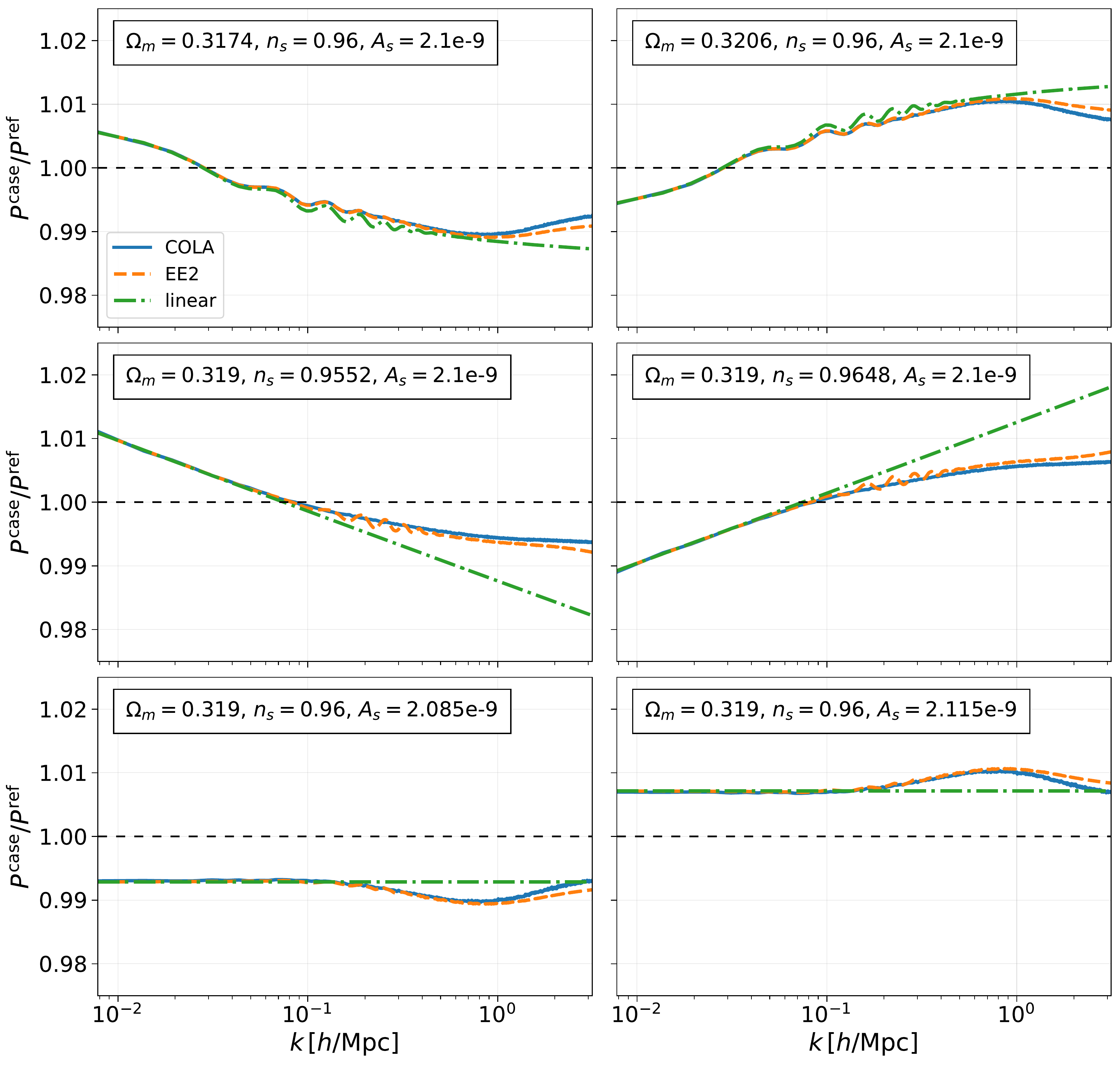} 
\caption{Non-linear and linear response functions computed using COLA, EE2 and \hiclass, for small variations of the cosmological parameters.}
\label{fig:cosmo_response_COLA_EE2_lin_small_var}
\end{figure}

In this Appendix, we show the results for small variations of cosmological parameters between COLA and EE2. Our choice of small variations represent increasing and decreasing $0.5\%$ of the reference value of $\Omega_{\rm m}$, $A_{\rm s}$ and $n_{\rm s}$. This small change in the parameters has smaller effects on the linear and non-linear response functions, as opposed to the large variation cases considered in Figure~\ref{fig:cosmo_response_COLA_EE2_lin}. Their impact on the matter power spectrum is shown in Figure~\ref{fig:cosmo_response_COLA_EE2_lin_small_var}. It is important to check that COLA can reproduce the response function with much better accuracy in the case. 

\begin{figure}[h] 
\centering
\includegraphics[width=1.0\textwidth]{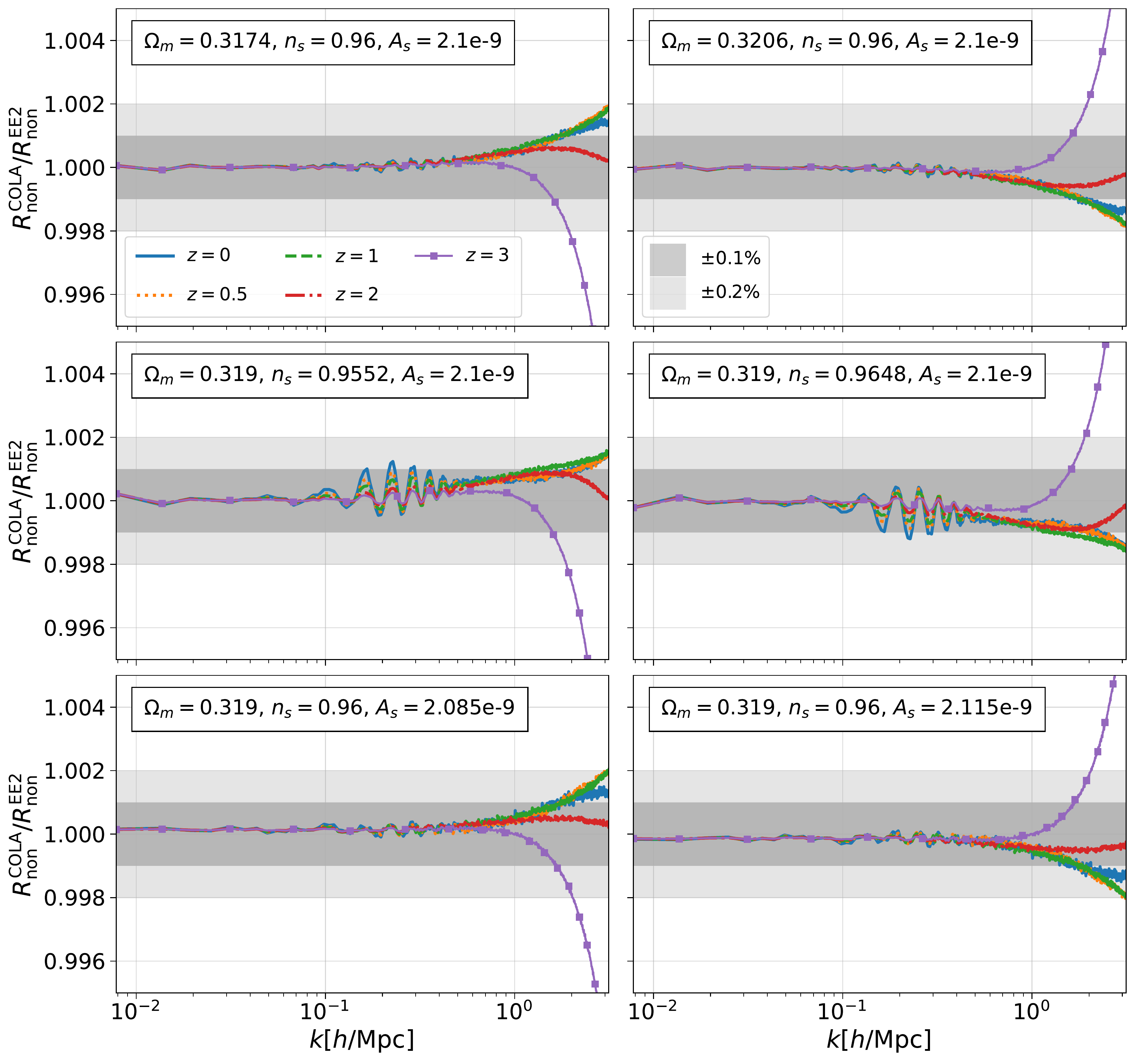} 
\caption{The same ratio as in Figure~\ref{fig:mnu0_COLA_EE2_ratio_Rs}, but for small variations of the cosmological parameters. 
}
\label{fig:Rnon_COLAvsEE2_SmallVar}
\end{figure}

\begin{figure}[h] 
\centering
\includegraphics[width=1.0\textwidth]{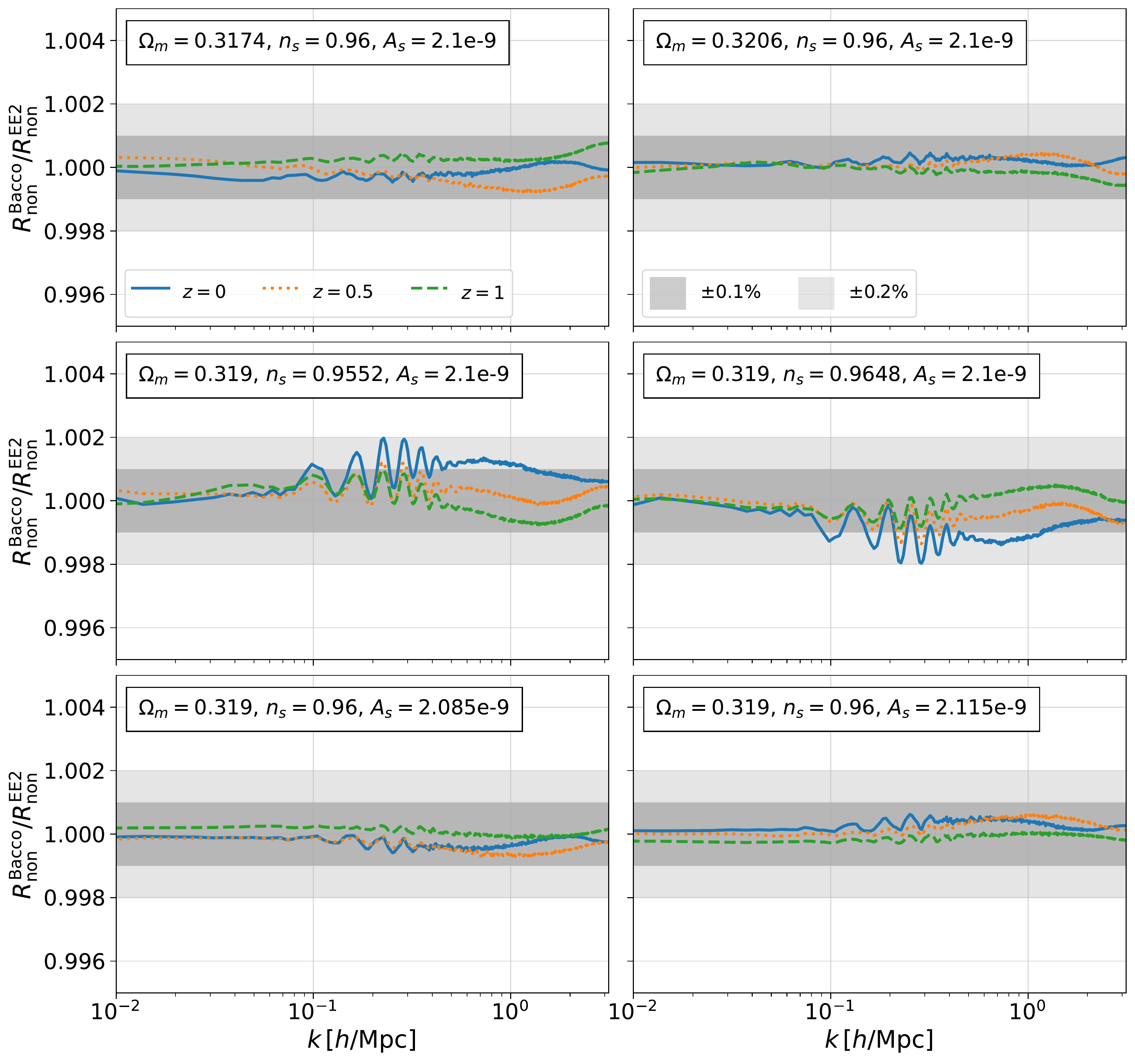} 
\caption{The same ratio as in Figure~\ref{fig:Rnon_BaccovsEE2_LargeVar}, but for small variations of the cosmological parameters. 
}
\label{fig:Rnon_BaccovsEE2_SmallVar}
\end{figure}

In Figures~\ref{fig:Rnon_COLAvsEE2_SmallVar} and \ref{fig:Rnon_BaccovsEE2_SmallVar} we show the impact on the ratio of the non-linear response of this change for COLA and EE2, and for Bacco and EE2, respectively. We see that in all cases COLA agrees with these emulators well within the $0.1\%$ threshold for up until $k=1$ $h$/Mpc.

\section{Convergence tests}
\label{sec:AppC}
To check the convergence of our result for the response function $R_{\rm non}$, we ran simulations with a large number of time steps as well as higher mass and force resolutions as a benchmark. The COLA method effectively converges to a standard Particle-Mesh (PM) N-body method with a large number of time steps. 

For this convergence test, we reduced the box size to $512$ Mpc$/h$. We then increased the number of time-steps for these simulations to $150$. The choice of $150$ time-steps is motivated by the number of steps used by MG-GLAM~\cite{ruan_2021}, another modified gravity PM based N-Body code. These simulations have 3 times better time resolution, 8 times better mass resolution and 3 times better force resolution than COLA simulation used in the analysis in this paper. We ran these higher resolution PM simulations for the ``cB10'' model discussed in the paper and a corresponding LCDM model. The ``cB10'' model has a 30\% enhancement of the linear power spectrum compared with the LCDM model, and the amplitude of $R_{\rm non}/R_{\rm lin}$ reaches 1.1, corresponding to the large variation of the cosmological parameters studied in section 3. We call these high-resolution simulations PM simulations. 

\begin{figure}[h] 
    \centering
    \includegraphics[width=0.8\textwidth]{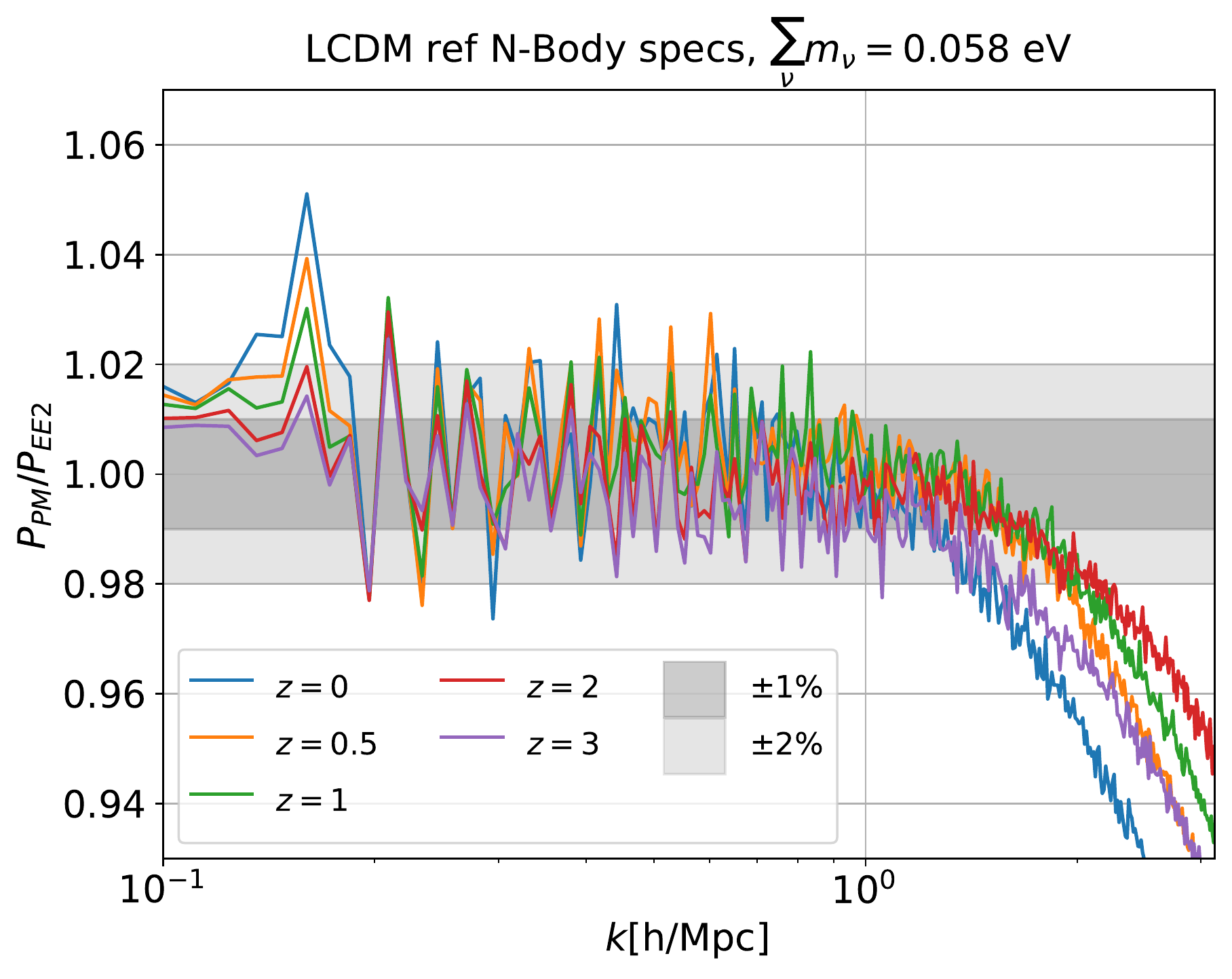}
    \caption{Ratio between the non-linear power spectra computed with the high-resolution PM simulation and the spectrum computed using EE2 for the reference LCDM cosmology.}
    \label{fig:plot2}
    \end{figure}

We checked that $P_{\rm non}(k)$ in the LCDM model from this PM simulation agrees with the Euclid emulator 2 prediction at $1 \%$ level at $k=1$ $h/$Mpc at all redshifts used in the paper as shown in Figure~\ref{fig:plot2}. This is an improvement compared with the result shown in Figure 11 for COLA simulations, where we have a $4\%$ agreement using the settings shown in Table 1.

\begin{figure}[ht] 
    \centering
    \includegraphics[width=0.9\textwidth]{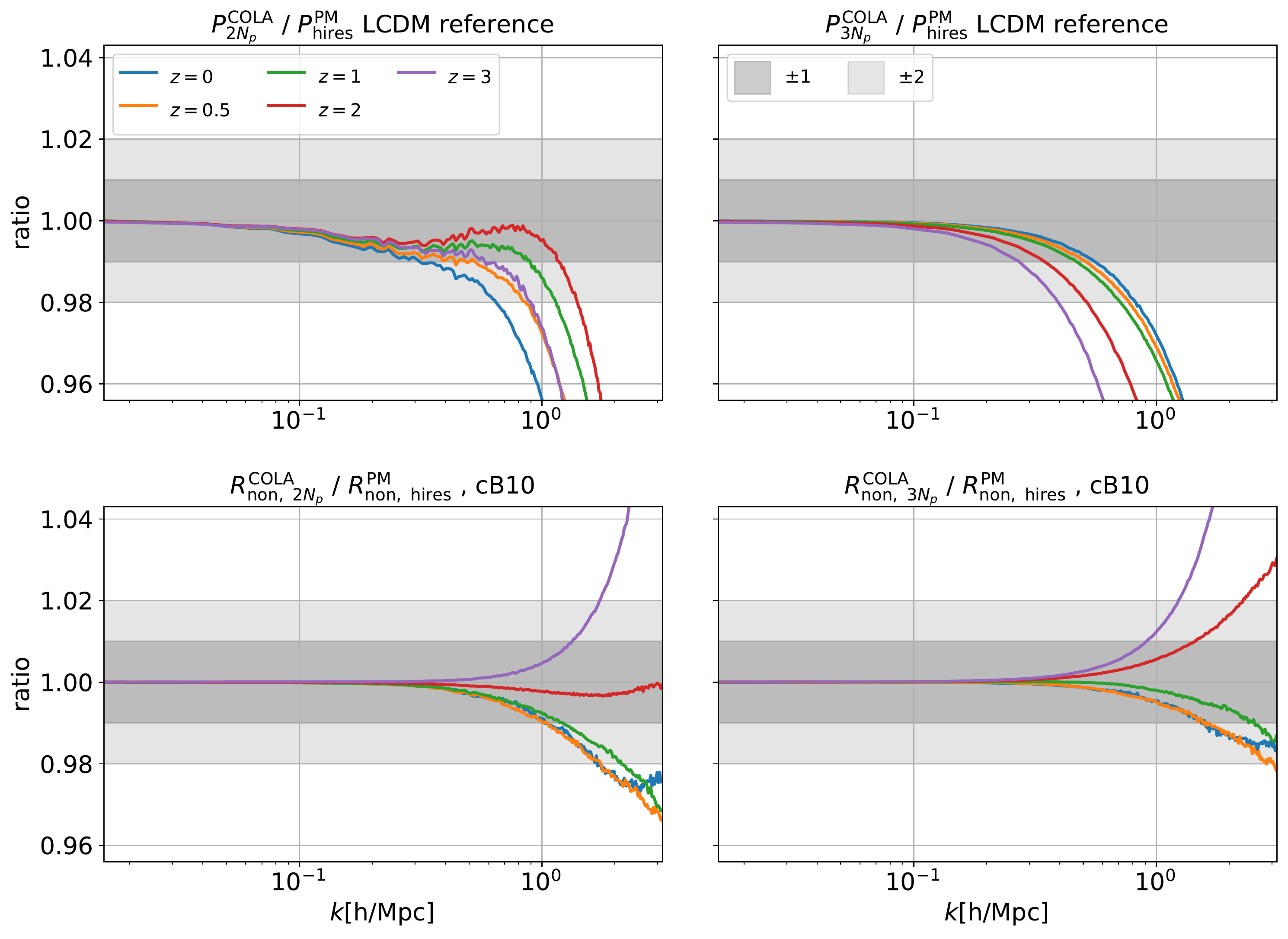}
    \caption{\textbf{Top row:} LCDM Matter power spectra ratio computed with COLA using different mesh resolutions, two times the number of particles per dimension (left) and three times the number of particles per dimension (right), in both cases the reference spectra is given by high-resolution PM simulations. \textbf{Bottom row:} cB10 non-linear response ratio computed with COLA using different mesh resolutions, two times the number of particles per dimension (left) and three times the number of particles per dimension (right), in both cases the reference response function is given by high-resolution PM simulations.}
    \label{fig:plot1}
    \end{figure}

In order to investigate the dependency of $R_{\mathrm{non}}$ and $P_{\mathrm{non}}$ on COLA settings, we ran the other 4 simulations, where we decreased the number of particles to 512 particles to have the same mass resolution as the one used in the paper, varied the mesh resolution to two and three times the number of particles per dimension and used the same time-steps as described in Table 6. Our results are summarized in Figure~\ref{fig:plot1}. The "COLA 2$N_{p}$" simulations have the same settings as those used in this paper.

\begin{table}[h]
    \begin{center} 
        \begin{tabular}{lcccc} 
            \toprule
            simulations & Volume  &
            Number of particles &
            Number of grids &
            Number of time steps \\ 
            \midrule
            PM & $512^3$ 
            & $1024^3$ & $(3N_p)^3$= $3072^3$ & 150 \\
            COLA 3$N_{p}$ & $512^3$ 
            & $512^3$ & $(3N_p)^3$=$1536^3$ & 50 \\
            COLA 2$N_{p}$ & $512^3$ 
            & $512^3$ & $(2N_p)^3$=$1024^3$ & 50 \\
            \bottomrule						
        \end{tabular}
    \end{center}
    \caption{Simulations for convergence tests. The volume is in the unit of {\rm Mpc}$^{3}$/$h^{3}$. Simulations in the last row have the same settings as COLA simulations used in the main text.}
    \label{table:convergence} 
\end{table}

We see that at higher redshifts the power spectrum in these COLA simulations suffers from the effect of low mass resolution, and increasing $N_{\mathrm{mesh}}$ does not improve the agreement with high-resolution results. The mass resolution effect seen in simulations with $N_{\mathrm{mesh}}=3 N_p$ is consistent with what was found in \cite{EE2}. 
We can also check that $N_{\mathrm{mesh}}=2 N_p$ simulations already give us a $1\%$ agreement in the response function, as we claim in the paper, and increasing the force resolution pushes the agreement slightly beyond $k=1$ $h/$Mpc at late times. However, we still have the problem of low mass resolution at higher redshifts. These plots show that the number of time-steps does not affect much $R_{\mathrm{non}}$, while the mass resolution effect is under control, as the effect stays below $1\%$ at $k=1$ $h/$Mpc in the response function.
This result is consistent with what we found in comparison with EE2 in LCDM as shown in section 3, and verifies the robustness of our prediction for $R_{\mathrm{non}}$ up to $k =1$ $h/$Mpc in this beyond LCDM model.  


\end{document}